\documentclass[twocolumn]{aastex631} 

\newcommand{\jwst}{{\it JWST}}
\newcommand{\hst}{{\it HST}}
\newcommand{\lya}{Ly$\alpha$}
\newcommand{\lephare}{\textsc{LePHARE}}
\newcommand{\ebv}{E(B-V)}
\newcommand{\mstar}{M$_\star$}

\defcitealias{Calzetti+00}{C00}
\defcitealias{Bruzual+03}{BC03}
\defcitealias{Schlafly+11}{SF11}

\usepackage{amsmath}

\begin{document}

\title{MIDIS: \jwst\ NIRCam and MIRI unveil the stellar population properties of \lya-emitters and Lyman-Break galaxies at $z \simeq 3 - 7$}

\correspondingauthor{Edoardo Iani}
\email{iani@astro.rug.nl, E.Iani@rug.nl}

\author[0000-0001-8386-3546]{Edoardo Iani}
\affiliation{Kapteyn Astronomical Institute, University of Groningen, P.O. Box 800, 9700AV Groningen, The Netherlands}

\author[0000-0001-8183-1460]{Karina I. Caputi}
\affiliation{Kapteyn Astronomical Institute, University of Groningen, P.O. Box 800, 9700AV Groningen, The Netherlands}
\affiliation{Cosmic Dawn Center (DAWN), Denmark}

\author[0000-0002-5104-8245]{Pierluigi Rinaldi}
\affiliation{Kapteyn Astronomical Institute, University of Groningen, P.O. Box 800, 9700AV Groningen, The Netherlands}

\author[0000-0002-8053-8040]{Marianna Annunziatella}
\affiliation{Centro de Astrobiolog\'ia (CAB), CSIC-INTA, Ctra. de Ajalvir km 4, Torrej\'on de Ardoz, E-28850, Madrid, Spain}
\affiliation{INAF-Osservatorio Astronomico di Capodimonte, Via Moiariello 16, 80131 Napoli, Italy}

\author[0000-0002-3952-8588]{Leindert A. Boogaard}
\affiliation{Max Planck Institute for Astronomy, K\"onigstuhl 17, 69117 Heidelberg, Germany}

\author[0000-0002-3005-1349]{G\"oran \"Ostlin}
\affiliation{Department of Astronomy, Stockholm University, Oscar Klein Centre, AlbaNova University Centre, 106 91 Stockholm, Sweden}


\author[0000-0001-6820-0015]{Luca Costantin}
\affiliation{Centro de Astrobiolog\'ia (CAB), CSIC-INTA, Ctra. de Ajalvir km 4, Torrej\'on de Ardoz, E-28850, Madrid, Spain}

\author[0000-0001-9885-4589]{Steven Gillman}
\affiliation{Cosmic Dawn Center (DAWN), Denmark}
\affiliation{DTU Space, Technical University of Denmark, Elektrovej, Building 328, 2800, Kgs. Lyngby, Denmark}

\author[0000-0003-4528-5639]{Pablo G. P\'erez-Gonz\'alez}
\affiliation{Centro de Astrobiolog\'ia (CAB), CSIC-INTA, Ctra. de Ajalvir km 4, Torrej\'on de Ardoz, E-28850, Madrid, Spain}


\author[0000-0002-9090-4227]{Luis Colina}
\affiliation{Centro de Astrobiolog\'ia (CAB), CSIC-INTA, Ctra. de Ajalvir km 4, Torrej\'on de Ardoz, E-28850, Madrid, Spain}

\author[0000-0002-2554-1837]{Thomas R. Greve}
\affiliation{Cosmic Dawn Center (DAWN), Denmark}
\affiliation{DTU Space, Technical University of Denmark, Elektrovej, Building 328, 2800, Kgs. Lyngby, Denmark}
\affiliation{Department of Physics and Astronomy, University College London, Gower Street, WC1E 6BT, UK}

\author[0000-0001-7416-7936]{Gillian Wright}
\affiliation{UK Astronomy Technology Centre, Royal Observatory Edinburgh, Blackford Hill, Edinburgh EH9 3HJ, UK}


\author[0000-0001-6794-2519]{Almudena Alonso-Herrero}
\affiliation{Centro de Astrobiolog\'ia (CAB), CSIC-INTA, Camino Bajo del Castillo s/n, E-28692 Villanueva de la Ca\~nada, Madrid, Spain}

\author[0000-0002-7093-1877]{Javier \'Alvarez-M\'arquez}
\affiliation{Centro de Astrobiolog\'ia (CAB), CSIC-INTA, Ctra. de Ajalvir km 4, Torrej\'on de Ardoz, E-28850, Madrid, Spain}

\author[0000-0001-8068-0891]{Arjan Bik}
\affiliation{Department of Astronomy, Stockholm University, Oscar Klein Centre, AlbaNova University Centre, 106 91 Stockholm, Sweden} 

\author[0000-0001-8582-7012]{Sarah E. I. Bosman}
\affiliation{Institute for Theoretical Physics, Heidelberg University, Philosophenweg 12, D–69120, Heidelberg, Germany}
\affiliation{Max-Planck-Institut f\"ur Astronomie, K\"onigstuhl 17, 69117 Heidelberg, Germany}

\author[0000-0003-2119-277X]{Alejandro Crespo G\'omez}
\affiliation{Centro de Astrobiolog\'ia (CAB), CSIC-INTA, Ctra. de Ajalvir km 4, Torrej\'on de Ardoz, E-28850, Madrid, Spain}

\author{Andreas Eckart}
\affiliation{Physikalisches Institut der Universit\"at zu K\"oln, Z\"ulpicher Str. 77, 50937 K\"oln, Germany}

\author[0000-0002-4571-2306]{Jens Hjorth}
\affiliation{DARK, Niels Bohr Institute, University of Copenhagen, Jagtvej 128, 2200 Copenhagen, Denmark}

\author[0000-0002-2624-1641]{Iris Jermann}
\affiliation{Cosmic Dawn Center (DAWN), Denmark}
\affiliation{DTU Space, Technical University of Denmark, Elektrovej, Building 328, 2800, Kgs. Lyngby, Denmark}

\author[0000-0002-0690-8824]{Alvaro Labiano}
\affiliation{Telespazio UK for the European Space Agency, ESAC, Camino Bajo del Castillo s/n, E-28692 Villanueva de la Ca\~nada, Madrid, Spain}
\affiliation{Centro de Astrobiolog\'ia (CAB), CSIC-INTA, Camino Bajo del Castillo s/n, E-28692 Villanueva de la Ca\~nada, Madrid, Spain}

\author[0000-0001-5710-8395]{Danial Langeroodi}
\affiliation{DARK, Niels Bohr Institute, University of Copenhagen, Jagtvej 128, 2200 Copenhagen, Denmark}

\author[0000-0003-0470-8754]{Jens Melinder}
\affiliation{Department of Astronomy, Stockholm University, Oscar Klein Centre, AlbaNova University Centre, 106 91 Stockholm, Sweden}

\author[0000-0002-3305-9901]{Thibaud Moutard}
\affiliation{Aix Marseille Universit\'e, CNRS, LAM (Laboratoire d’Astrophysique de Marseille) UMR 7326, F-13388, Marseille, France}

\author{Florian Pei{\ss}ker}
\affiliation{Physikalisches Institut der Universit\"at zu K\"oln, Z\"ulpicher Str. 77, 50937 K\"oln, Germany}

\author[0000-0002-0932-4330]{John P. Pye}
\affiliation{School of Physics \& Astronomy, Space Research Centre, Space Park Leicester, University of Leicester, 92 Corporation Road, Leicester LE4 5SP, UK}

\author[0009-0003-6128-2347]{Tuomo V. Tikkanen}
\affiliation{School of Physics \& Astronomy, Space Research Centre, Space Park Leicester, University of Leicester, 92 Corporation Road, Leicester LE4 5SP, UK}

\author[0000-0001-5434-5942]{Paul P. van der Werf}
\affiliation{Leiden Observatory, Leiden University, P.O. Box 9513, 2300 RA Leiden, The Netherlands}

\author[0000-0003-4793-7880]{Fabian Walter}
\affiliation{Max Planck Institute for Astronomy, K\"onigstuhl 17, 69117 Heidelberg, Germany}
\affiliation{National Radio Astronomy Observatory, Pete V. Domenici Array Science Center, P.O. Box O, Socorro, NM 87801, USA}


\author[0000-0002-1493-300X]{Thomas K. Henning}
\affiliation{Max-Planck-Institut f\"ur Astronomie, K\"onigstuhl 17, 69117 Heidelberg, Germany}

\author{Pierre-Olivier Lagage}
\affiliation{Universit\'e Paris-Saclay, Universit\'e Paris Cit\'e, CEA, CNRS, AIM, F-91191 Gif-sur-Yvette, France}

\author[0000-0001-7591-1907]{Ewine F. van Dishoeck}
\affiliation{Leiden Observatory, Leiden University, P.O. Box 9513, 2300 RA Leiden, The Netherlands}
\affiliation{Max-Planck Institut f\"ur Extraterrestrische Physik (MPE), Giessenbachstr. 1, 85748,
Garching, Germany}




\begin{abstract}
We study the stellar population properties of 182 spectroscopically-confirmed (MUSE/VLT) Lyman-$\alpha$ emitters (LAEs) and 450 photometrically-selected Lyman-Break galaxies (LBGs) at $z = 2.8 - 6.7$ in the Hubble eXtreme Deep Field (XDF). 
Leveraging the combined power of \hst\ and \jwst\ NIRCam and MIRI observations, we analyse their rest-frame UV-through-near-IR spectral energy distributions (SEDs) with MIRI playing a crucial role in robustly assessing the LAE's stellar mass and ages. 
Our LAEs are low-mass objects ($\rm \log_{10}(M_\star~[M_\odot]) \simeq 7.5$), with little or no dust extinction ($\rm E(B-V) \simeq 0.1$) and a blue UV continuum slope ($\beta \simeq -2.2$).
While 75\% of our LAEs are young ($<$~100 Myr), the remaining 25\% have significantly older stellar populations ($\geq 100$~Myr). 
These old LAEs are statistically more massive, less extinct and have lower specific star formation rate (sSFR) compared to young LAEs. Besides, they populate the M$_\star$ -- SFR plane along the main-sequence (MS) of star-forming galaxies, while young LAEs populate the starburst region. 
The comparison between the LAEs properties to those of a stellar-mass matched sample of LBGs shows no statistical difference between these objects, except for the LBGs redder UV continuum slope and marginally larger E(B-V) values. 
Interestingly, 48\% of the LBGs have ages $<$~10 Myr and are classified as starbursts, but lack detectable \lya\ emission. This is likely due to HI resonant scattering and/or selective dust extinction.
Overall, we find that \jwst\ observations are crucial in determining the properties of LAEs and shedding light on their comparison with LBGs.

\end{abstract}



\section{Introduction} \label{sec:intro}

The Lyman-$\alpha$ \cite[\lya,][]{Lyman+06} is the brightest emission line produced by hydrogen electronic transitions, having an energy of 10.2~eV and a wavelength of 1215.67 \AA.
Given that approximately 74\% of the Universe baryonic matter is thought to consist of hydrogen atoms \cite[e.g.][]{Croswell+96, Carroll+06}, this ultraviolet (UV) transition emerges as a highly effective tool for detecting galaxies due to its brightness. 
This is particularly advantageous at intermediate to high redshifts (around $z\gtrsim 2-3$) because the \lya\ UV rest-frame wavelength undergoes a cosmological redshift, shifting it into the optical and near-infrared (NIR) regions of the electromagnetic spectrum. This shift enables convenient observations using both ground-based and space-based facilities.

Galaxies detected thanks to their \lya\ emission are generally referred to as \lya-{\it emitters} \cite[LAEs, e.g.][]{Ouchi+20}.
The \lya\ detection can be direct, whenever based on spectroscopic datasets \cite[e.g.][]{Herenz+17, Herenz+19, Claeyssens+22, Bacon+23}, or indirect, when inferred from narrow-band photometry \cite[e.g.][]{Ouchi+08, Ouchi+10, Ouchi+18, Ota+10, Ota+17, Shibuya+12, Konno+14, Santos+16, Zheng+17, ArrabalHaro+18}.
In the last decades, studies have thoroughly investigated the rest-frame UV physical properties of intermediate/high-redshift LAEs mainly leveraging on the available optical spectroscopy and imaging. 
This allowed astronomers to find that, in the absence of an active galactic nucleus (AGN), LAEs are low-mass star-forming galaxies (SFGs) (stellar mass $\rm M_\star\lesssim 10^{8-9} ~M_\odot$), with young stellar populations (stellar ages $\simeq 10$ Myr) and star formation rates ${\rm SFR} \simeq 1-10~ {\rm M}_\odot/{\rm yr}$ \cite[e.g.][]{Nakajima+12, Hagen+14, Hagen+16}.
Besides, LAEs were found to be dust-poor galaxies with stellar and nebular colour extinction values ${\rm E(B-V)} \simeq 0-0.2$~\cite[][]{Ono+10, Kojima+17}, and a sub-solar gas-phase metallicity ${\rm Z}\simeq 0.1 - 0.5~{\rm Z}_\odot$ \cite[derived from both strong lines and direct electron temperature T$_e$ methods, e.g.][]{Finkelstein+11, Nakajima+12, Trainor+16, Kojima+17}. 

The coarser spatial resolution and sensitivity of the available near- and mid-infrared (MIR) instrumentation at the time (e.g. \textit{Spitzer}) strongly limited the analysis of the LAEs rest-frame optical/NIR properties at $z \gtrsim 2 - 3$.
Since the rest-frame UV emission of galaxies is known to trace the youngest, brightest and less obscured stellar populations within a galaxy, our overall general interpretation of LAEs properties could be possibly biased.
In this regard, a few observational studies highlighted that some LAEs appear to have an underlying stellar population older than what is generally found, with ages significantly above 100 Myr \cite[][]{Lai+08, Finkelstein+09, Pentericci+09, Nilsson+09, Rosani+20}, thus suggesting the existence of two classes of LAEs already at intermediate/high-redshift.
Theoretical studies advocate that these two classes of LAEs are the consequence of catching these galaxies in different stages of their evolution \cite[e.g.][]{Shimizu+10}: while LAEs hosting a young stellar population ($< 100$~Myr) would be early coeval starbursts due to the contemporary accretion of sub-haloes in a young small parent halo (\textit{primeval galaxies}), LAEs with an underlying old stellar age ($> 100$~Myr) could be delayed starbursts triggered by later sub-halo accretion onto evolved haloes (\textit{rejuvenation}).

Contextually, strong debates took place in understanding if LAEs have or have not different physical properties with respect to SFGs at similar redshifts but that do not display \lya\ emission.
In the recent literature \cite[e.g.][]{Dayal+12, deLaVieuville+20, ArrabalHaro+20}, these galaxies are often referred to as Lyman-Break Galaxies (LBGs) since they are typically found on the basis of broad-band photometry via the Lyman-Break technique \cite[][]{Steidel+96} and spectral energy distribution (SED) fitting methods.
On this topic, in the last two decades, both observational and theoretical studies reached different conclusions. While some works have found no substantial difference between LAEs and LBGs \cite[e.g.][]{Dayal+12}, others found LBGs to have higher stellar-mass with less rapid star formation compared to LAEs \cite[e.g.][]{Giavalisco+02, Gawiser+06}.

Today, thanks to the Near-Infrared Camera \cite[NIRCam;][]{Rieke+05} and  Mid-Infrared Instrument \cite[MIRI;][]{Rieke+15, Wright+15} onboard the {\it James Webb Space Telescope} (\jwst), we can finally address these open questions by inquiring into the nature of LAEs and LBGs at $z \gtrsim 3$ and understand their similarities and/or differences. 
In this sense, the unprecedented spatial resolution and depth reached by MIRI imaging at wavelengths $\gtrsim 5.6~\mu$m plays a crucial role for the detection of these sources optical/near-infrared rest-frame emission \citep{Wright+23}. 
In particular, between $z = 2.8 - 6.7$, the imaging at 5.6~$\mu$m allows for the characterisation of the stellar optical emission in the range $\lambda \simeq 0.7 - 1.5~\mu$m, i.e. red-wards of the hydrogen Balmer transition of H$\alpha$ at 6562.8~\AA.
By probing a region of the optical spectrum not affected by strong emission lines, it ensures a much more robust determination of the galaxy physical properties \cite[e.g.][]{Papovich+23}.

In this paper, we present our findings on the physical properties of the stellar populations of a sample of 182 LAEs at redshift $z=2.8-6.7$ identified in the Hubble eXtreme Deep Field \cite[XDF,][]{Illingworth+13} with the Multi Unit Spectroscopic Explorer \cite[MUSE,][]{Bacon+10} at the Very Large Telescope (VLT).
This analysis takes advantage of the synergy between archival \hst\ and the latest \jwst\ observations to constrain the SED of these sources.
In addition, we present a comparison of the properties of our sample of LAEs to those of a sample of 450 photometrically-selected LBGs in the same field and redshift range.

Throughout this paper, we consider a flat $\Lambda$-CDM cosmology with $\rm H_0=70 \,{\rm km \, s^{-1} Mpc^{-1}}$, $\rm \Omega_M=0.3$ and $\rm \Omega_\Lambda=0.7$. 
All magnitudes are total and refer to the AB system \citep{Oke+83}. 
Finally, we assume a \citet{Chabrier+03} initial mass function (IMF).

\begin{deluxetable*}{lcccccc}[t]
    \tablenum{1}
    \tablecaption{Dataset properties.}
    \tablewidth{0pt}
    \tablehead{
    \colhead{Instrument}   & \colhead{Filter} & \colhead{ZPT$_{\rm AB}$} & \colhead{$f_{\rm ext}$} & \colhead{$f_{\rm aper}$} & \colhead{$t_{\rm exp}$} & \colhead{Program}  \\
                           &        & \colhead{[mag]}          &               &                & \colhead{[ks]}          &              }
    \startdata
    \textit{HST}/WFC3-UVIS & F225W  & 24.04          & 1.0615        & 1.2477         & 89.58         & \citealt{Whitaker+19} \\                                             
    \textit{HST}/WFC3-UVIS & F275W  & 24.13          & 1.0469        & 1.2203         & 207.22        & \citealt{Whitaker+19} \\                                             
    \textit{HST}/WFC3-UVIS & F336W  & 24.67          & 1.0379        & 1.1833         & 216.15        & \citealt{Whitaker+19} \\                                             
    \textit{HST}/ACS       & F435W  & 25.68          & 1.0308        & 1.1587         & 526.95        & \citealt{Whitaker+19} \\                                             
    \textit{HST}/ACS       & F606W  & 26.51          & 1.0209        & 1.1547         & 537.94        & \citealt{Whitaker+19} \\                                             
    \textit{HST}/ACS       & F775W  & 25.69          & 1.0137        & 1.1655         & 770.19        & \citealt{Whitaker+19} \\                                             
    \textit{HST}/ACS       & F814W  & 25.94          & 1.0128        & 1.1723         & 840.21        & \citealt{Whitaker+19} \\                                             
    \textit{HST}/ACS       & F850LP & 24.87          & 1.0104        & 1.2346         & 1495.09       & \citealt{Whitaker+19} \\                                             
    \textit{JWST}/NIRCam   & F090W  & 26.80          & 1.0104        & 1.1601         & 53.50         & \textit{JADES}        \\                                     
    \textit{HST}/WFC3-IR   & F098M  & 27.19          & 1.0089        & 1.2225         & 484.73        & \citealt{Whitaker+19} \\                                             
    \textit{HST}/WFC3-IR   & F105W  & 27.78          & 1.0080        & 1.2330         & 60.50         & \citealt{Whitaker+19} \\                                             
    \textit{JWST}/NIRCam   & F115W  & 26.91          & 1.0068        & 1.1587         & 440.76        & \textit{JADES}        \\                                     
    \textit{HST}/WFC3-IR   & F125W  & 27.74          & 1.0059        & 1.2674         & 440.76        & \citealt{Whitaker+19} \\                                             
    \textit{HST}/WFC3-IR   & F140W  & 27.96          & 1.0048        & 1.3089         & 115.59        & \citealt{Whitaker+19} \\                                             
    \textit{JWST}/NIRCam   & F150W  & 27.32          & 1.0042        & 1.1648         & 35.70         & \textit{JADES}        \\                                     
    \textit{HST}/WFC3-IR   & F160W  & 27.45          & 1.0040        & 1.3587         & 596.28        & \citealt{Whitaker+19} \\                                            
    \textit{JWST}/NIRCam   & F182M  & 26.73          & 1.0029        & 1.1710         & 32.29         & \textit{FRESCO $+$ JEMS}         \\                                    
    \textit{JWST}/NIRCam   & F200W  & 27.36          & 1.0026        & 1.1716         & 24.70         & \textit{JADES}        \\                                     
    \textit{JWST}/NIRCam   & F210M  & 26.48          & 1.0024        & 1.1723         & 31.35         & \textit{FRESCO $+$ JEMS}         \\                                    
    \textit{JWST}/NIRCam   & F277W  & 28.86          & 1.0017        & 1.1955         & 35.70         & \textit{JADES}        \\                                     
    \textit{JWST}/NIRCam   & F335M  & 28.10          & 1.0014        & 1.2369         & 24.70         & \textit{JADES}        \\                                     
    \textit{JWST}/NIRCam   & F356W  & 29.07          & 1.0013        & 1.2523         & 24.70         & \textit{JADES}        \\                                     
    \textit{JWST}/NIRCam   & F410M  & 28.24          & 1.0012        & 1.2945         & 35.70         & \textit{JADES}         \\                                    
    \textit{JWST}/NIRCam   & F430M  & 27.40          & 1.0011        & 1.3063         & 13.92         & \textit{JEMS}         \\                                    
    \textit{JWST}/NIRCam   & F444W  & 29.11          & 1.0011        & 1.3115         & 36.63         & \textit{FRESCO $+$ JADES}        \\                                     
    \textit{JWST}/NIRCam   & F460M  & 27.20          & 1.0011        & 1.3271         & 13.92         & \textit{JEMS}         \\                                    
    \textit{JWST}/NIRCam   & F480M  & 27.36          & 1.0010        & 1.3414         & 27.83         & \textit{JEMS}         \\                                    
    \textit{JWST}/MIRI     & F560W  & 28.68          & 1.0010        & 1.6603         & 148.84        & \textit{MIDIS}        \\                                    
    \enddata
    \tablecomments{In the above table, we report the main properties of our dataset. For the 28 filters adopted, we present the corresponding photometric zero-points (in AB magnitude) ZPT$_{\rm AB}$, the multiplicative factor $f_{\rm ext}$ to correct the fluxes for Galactic extinction, the multiplicative factor $f_{\rm aper}$ to correct the fluxes derived via aperture photometry for PSF effects, the total exposure time $t_{\rm exp}$ and, finally, the observations programs. 
    The reported ZPT$_{\rm AB}$ values are valid for images in electrons/s (both \hst\ and \jwst).
    The $f_{\rm ext}$ values were obtained assuming a colour excess $\rm E(B-V) = 0.008$\footnote{The Galactic extinction colour excess was obtained from the IRSA webpage (\url{https://irsa.ipac.caltech.edu/applications/DUST/}) and corresponds to the value at the celestial coordinates at the centre of the \jwst/MIRI F560W pointing.} \citepalias{Schlafly+11}. Following \citetalias{Schlafly+11}, for all filters with an effective wavelength $\lambda_{\rm eff} < 1.25 \mu$m, we derive $f_{\rm ext}$ applying the \cite{Fitzpatrick+99} extinction law, while at longer wavelengths ($1.25 \leq \lambda_{\rm eff} < 8 \mu$m),  we resort to the \cite{Indebetouw+05} law. The $f_{\rm aper}$ values were obtained considering circular apertures of 0.5 arcsec diameter. For the \hst\ filters, we adopt known aperture correction values (\url{https://www.stsci.edu/hst/instrumentation/}). 
    For the \jwst/NIRCam filters, we estimate the aperture corrections using the \textsc{WebbPSF} software. For the MIRI/F560W, we adopt the PSF reconstructed by \cite{Boogaard+23} that effectively models the phenomenon of internal diffraction occurring inside the MIRI detector at $\lesssim 10~\mu$m \cite[][]{Gaspar+21} and that is not currently reproduced by \textsc{WebbPSF}.}
    \label{tab:data_prop}
\end{deluxetable*}

\section{Data} \label{sec:style}
The XDF \citep[$\alpha_{J2000} = 3^h 32^m 38.5^s$, $\delta_{J2000} = -27^\circ 47' 00.0''$; ][]{Illingworth+13} is a small field of the sky with the deepest \textit{Hubble Space Telescope} (\textit{HST}) observations ever taken since this telescope started operations more than thirty years ago. This field has been the main window to study the early Universe before the \textit{JWST} advent, with numerous works scientifically exploiting its unique possibilities. 
Now, in the \textit{JWST}'s era, the \textit{HST} data in the XDF and surroundings are being enhanced with deep imaging and spectroscopy obtained with \textit{JWST/}NIRCam and MIRI, extending the wavelength coverage of high spatial-resolution observations to the near- and mid-infrared.
In addition to this rich photometric dataset, the XDF has been the target of extended spectroscopic campaigns including the observations with the MUSE at VLT.
These observations have provided thousands of spectra for sources up to $z \simeq 6.7$. 
In this Section, we briefly describe the dataset used for our study which includes MUSE spectroscopy, and imaging from both \hst\ (ACS and WFC3) and \jwst\ (NIRCam and MIRI)\footnote{
All the {\it HST} and {\it JWST} data used in this paper can be found in the Mikulski Archive for Space Telescopes (MAST): \dataset[10.17909/T91019]{http://dx.doi.org/10.17909/T91019}, \dataset[10.17909/8tdj-8n28]{http://dx.doi.org/10.17909/8tdj-8n28}, \dataset[10.17909/gdyc-7g80]{http://dx.doi.org/10.17909/gdyc-7g80}, \dataset[10.17909/fsc4-dt61]{http://dx.doi.org/10.17909/fsc4-dt61}, \dataset[10.17909/5txh-pj89]{http://dx.doi.org/10.17909/5txh-pj89}.}.

\subsection{VLT/MUSE}
The XDF has been extensively studied with the VLT/MUSE over the past nine years as part of the MOSAIC and UDF-10 fields (GTO programs 094.A-0289(B), 095.A-0010(A), 096.A-0045(A) and 096.A-0045(B), PI: R. Bacon), and the most recent MXDF observations (GTO Large Program 1101.A-0127, PI: R. Bacon). 
For more details about the observations related to the MOSAIC and UDF-10 fields, we refer to \cite{Bacon+17}, while we point the reader to \cite{Bacon+23} for a thorough explanation of the MXDF program. 

In brief, the three programs covered the XDF area with MUSE observations in Wide-Field Mode (WFM) with each single pointing covering an area of about 1 arcmin$^2$, a spectral wavelength range between 4700 - 9300 \AA\ and a spectral resolving power R that varies from 1770 (4800 \AA) to 3590 (9300 \AA).
While the UDF-10 and MOSAIC programs were carried out without the ground-layer adaptive optics (GLAO) mode of the VLT Adaptive Optics Facility (AOF) via the GALACSI adaptive optics module \cite[][]{Kolb+16, Madec+18}, the MXDF program made use of VLT’s AOF and GALACSI. 
With respect to non-AOF observations, the only change in the MUSE instrumental configuration is the notch filter that blocks the light in the 5800 - 5966\AA\ wavelength range which, otherwise, would be strongly contaminated by the bright light of the four sodium laser guide stars used by the AOF. 
In WFM the MUSE spatial sampling is $0.2''\times 0.2''$, while the spatial resolution varies significantly between programs going from a median value of $\simeq 0.8$ arcsec (MOSAIC and UDF-10 programs) down to $0.4$ arcsec in the case of MXDF observations.
Within the XDF area, the observations can reach a maximum depth of more than 140h but the depth is not homogeneous.

\cite{Bacon+23} provide fully reduced MUSE datacubes for the MOSAIC + UDF-10 and MXDF programs, as well as a catalogue of detected sources and corresponding spectroscopic redshifts.\footnote{The MUSE cubes and catalogue of sources can be obtained at \url{https://amused.univ-lyon1.fr/project/UDF/HUDF/}.}

\subsection{JWST/NIRCam}
We utilise recent NIRCam images collected as part of the General Observers (GO) program \textit{JWST Extragalactic Medium-band Survey} (\textit{JEMS}, PID: 1963; PIs: C. C. Williams, S. Tacchella, M. Maseda) covering the HUDF. 
These observations were carried out in 5 medium bands (F182M, F210M, F430M, F460M, and F480M), with 7.8 hours of integration time dedicated to F182M, F210M, and F480M, and 3.8 hours for F430M and F460M \citep{Williams+23}. 

In addition, we use imaging data obtained as part of the GO program \textit{The First Reionization Epoch Spectroscopic COmplete Survey} (\textit{FRESCO}, PID: 1895; PI: P. Oesch) to complement the \textit{JEMS} dataset. \textit{FRESCO} provides additional imaging in the F182M and F210M filters as well as at F444W \citep{Oesch+23}.

We process all the NIRCam images from \textit{JEMS} and \textit{FRESCO} using a modified version of the official \jwst\ pipeline (version 1.8.2\footnote{CRDS context \texttt{jwst\_1018.pmap}}), which includes several procedures to minimise the impact of various image artefacts, such as {\it snowballs}, \textit{1/f} noise, {\it wisps}, and residual cosmic rays \cite[e.g.][]{Rinaldi+23a, Perez-Gonzalez+23}. 
After reducing the images, we drizzle them to a pixel scale of 0.03 arcsec/pixel and align them to the Hubble Legacy Fields (HLF) catalogue by \cite{Whitaker+19}. 

Finally, we complement the \textit{JEMS} and \textit{FRESCO} observations with the recently published  NIRCam data from the GO program \textit{The JWST Advanced Deep Extragalactic Survey} (\textit{JADES}, PIDs: 1180, 1210; P.I.: D. Eisenstein, N. Luetzgendorf). 
The \textit{JADES} dataset adds deep-imaging in eight \jwst/NIRCam bands F090W, F115W, F150W, F200W, F277W, F335M, F356W, F410M, and significantly increases the depth in the F444W filter \citep{Eisenstein+23}.
For our study, we download the fully-reduced publicly released \textit{JADES} observations\footnote{The \textit{JADES} fully-reduced images are available at \url{https://archive.stsci.edu/hlsp/jades\#section-268de08a-1ff5-430e-adfe-846e6b933f3b}.} \citep{Rieke+23} from the Mikulski Archive for Space Telescopes (MAST).
Although this imaging data underwent a distinct optimisation process (developed by the JADES collaboration using the \jwst\ pipeline) compared to our own processing, a visual inspection of the JADES final products in the XDF region did not uncover any artefacts or patterns that might affect the quality of photometric measurements in those bands. Consequently, we consider the quality of the results produced by both pipelines as comparable.
After matching their astrometry to the HLF catalogue, we resample the \textit{JADES} images to the same scale as the \textit{JEMS} and \textit{FRESCO} NIRCam observations (0.03 arcsec/pixel).

\subsection{JWST/MIRI}
We complement the NIRCam observations with MIRI $5.6\mu$m imaging from the \jwst\ Guaranteed Time Observations (GTO) program {\it The MIRI HUDF Deep Imaging Survey} (\textit{MIDIS}, PID: 1283, PI: G. \"Ostlin). 
The MIRI observations cover an area of about 4.7 arcmin$^2$ of the XDF for about 40 hours of total integration time and were carried out in December 2022 using the broad-band filter F560W. 
This is the deepest $5.6\mu$m image available to date \cite[e.g.][]{Rinaldi+23a, Boogaard+23}. 

Similarly to the NIRCam observations, we use a modified version of the official \jwst\ pipeline (version 1.8.4\footnote{CRDS context \texttt{jwst\_1014.pmap}}) to address strong patterns, such as vertical striping and background gradients, that affect the scientific quality of the images \cite[e.g.][]{Iani+22, Rodighiero+23}. 
We add extra steps at the end of stages 2 and 3 of the official \jwst\ pipeline to mitigate these issues and reduce the noise in the output image. 
For more details about the latest \textit{MIDIS} data collection, we refer the reader to \"Ostlin et al. ({\it in prep.}).

Finally, we register the astrometry of the final MIRI image to the HLF catalogue and drizzle it to the same pixel scale as the NIRCam images.

\subsection{HST}
We combine our \jwst\ observations with the \hst\ images of the XDF obtained from the Hubble Legacy Field GOODS-S \cite[HLF-GOODS-S\footnote{The \textit{HST} imaging is available at \url{https://archive.stsci.edu/prepds/hlf/}},][]{Whitaker+19}. 
The HLF-GOODS-S comprises 13 \textit{HST} bands, spanning a broad range of wavelengths from 0.2$\mu$m to 1.6$\mu$m including UV (WFC3/UVIS: F225W, F275W, F336W), optical (ACS/WFC: F435W, F606W, F775W, F814W, F850LP), and near-infrared (WFC3/IR: F098M, F105W, F125W, F140W, F160W) filters. 
For more detailed information on these observations, we refer the reader to \citet{Whitaker+19}.

\begin{figure*}[t!]
    \centering
    \includegraphics[width=\textwidth, trim={1cm 1cm 1cm 1cm}]{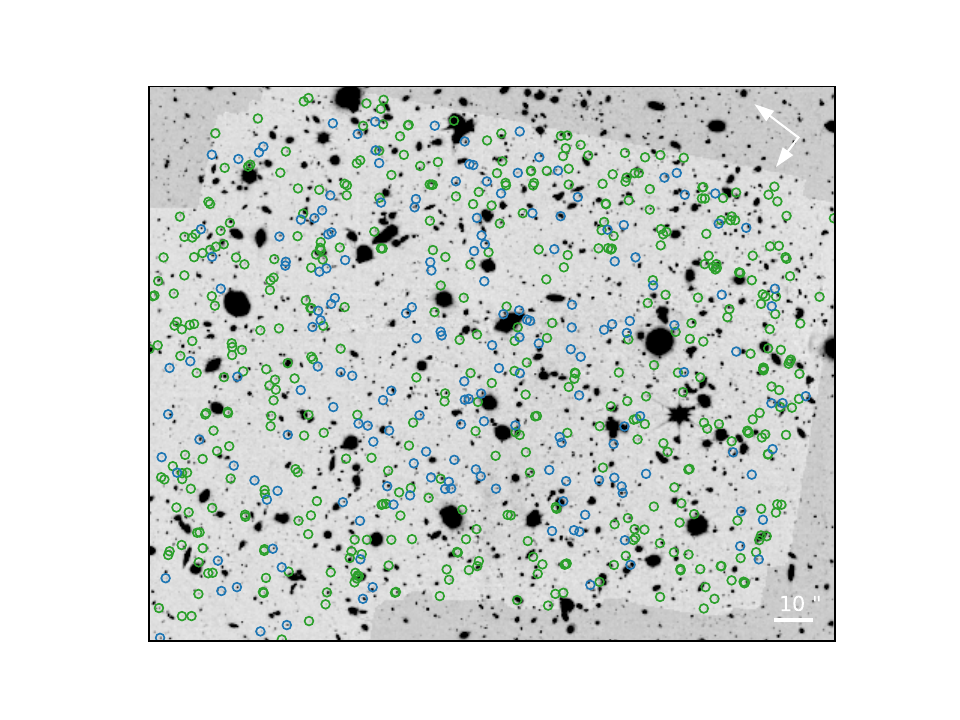}
    \caption{Stacked image of XDF obtained by combining the 28 \hst\ and \jwst\ filters available. We display LAEs in blue and LBGs in green. Circles are 1.2 arcsec in diameter. We highlight the XDF region covered by the \jwst/F560W MIRI image with a lighter background colour.}
    \label{fig:rgb_ima}
\end{figure*}

\section{Sample selection and multi-wavelength photometry} \label{sec:sample}
In the following Section, we describe the process we follow to select our sample of LAEs and find their UV/optical/near-IR counterpart in the \hst\ and \jwst\ imaging.
We also describe how we select a sample of galaxies in the same redshift range ($z\simeq 3-7$) as our sample of LAEs but that do not display \lya-emission (LBGs).

\subsection{Lyman-$\alpha$ Emitters in XDF} \label{subsec:laes_selection}
To define our sample of LAEs, we start from the publicly available catalogue by \cite{Bacon+23}. 
We select all the LAEs lying within the area covered by the MIRI observations.
We discard all sources reported with a quality flag on their spectroscopic redshift confirmation QF $= 1$ (indicative of a low confidence in the line identification) while keeping, instead, the LAEs with QF $=2$\footnote{According to the authors, a confidence level of 2 for LAEs indicates good confidence. The requirement is to have a \lya\ with SNR~$ > 5$ and a width and asymmetry compatible with \lya\ line shapes. A confidence level of 3 indicates high confidence, i.e., if there is no other line than \lya, they require an SNR~$> 7$ with the expected line shape: a pronounced red asymmetrical line profile and/or a blue bump or double-peaked line profile.}.
We retain the QF = 2 LAEs since they constitute 2/3 of the overall sample (QF~$=2,3$) and their removal would introduce significant biases towards brighter LAEs.
By doing so, we end up with an initial sample of 480 LAEs.

From a visual inspection, we find that 30 sources are in proximity ($\leq 1''$) of bright and extended galaxies at $z \leq 2.5$. 
The vicinity and brightness of these objects contaminate the photometry of the LAEs' counterparts.
Besides, the presence of massive foreground objects can lens and magnify our targets \cite[e.g.][]{Matthee+17}.
Hence, we remove these sources from our analysis. 

LAEs often host active galactic nuclei \cite[AGNs, e.g.][]{Ouchi+08}.
For this reason, we investigate if AGNs are hidden in our final catalogue of LAEs. In fact, due to their peculiar physical properties, the presence of AGNs in our sample could impact and contaminate our final results.
We look for AGNs in a two-fold way: we cross-match our final catalogue with X-ray observations in XDF, and investigate the presence of high-ionisation UV lines in the spectrum of our targets.
These two methodologies allow us to verify the presence of unobscured (Type I) and obscured (Type II) AGNs, respectively.
The publicly available X-ray catalogue based on the \textit{Chandra} observatory \cite[Chandra Source Catalog\footnote{The Chandra Source Catalog (CSC) is available at \url{https://cxc.cfa.harvard.edu/csc2/}}, version 2.0,][]{Evans+20} does not report any X-ray emitter at our targets coordinates nor in their closest vicinity ($\leq 1$ arcsec).
This is also confirmed from the comparison to the \textit{Chandra}-based catalogue by \cite{Luo+17} and the catalogue of AGNs in XDF based on \textit{XMM-Newton} observations by \cite{Ranalli+13}.  
As for the study of UV lines, the catalogue by \cite{Bacon+23} presents the equivalent width (EW) and flux of lines such as HeII$\lambda1640$ (HeII), and the  CIV$\lambda\lambda1548,1550$ (CIV) and CIII]$\lambda\lambda1907,1909$ (CIII]) doublets. 
Following \cite{Nakajima+18}, we can probe the presence of AGNs via the diagnostic diagrams CIV/CIII] vs. (CIII]+CIV)/HeII, EW(CIII]) vs. CIII]/HeII, and EW(CIV) vs. CIV/HeII.
According to \cite{Bacon+23}, however, only two sources in our sample have a detection of all these lines with a SNR~$\geq 3$.
Both objects lie in the locus of the diagrams populated by star-forming galaxies.
Similar results are obtained even when pushing the SNR down to $\simeq 2$. As an additional check, we extract the MUSE spectra of our sources, bring them to rest-frame and stack them. The final stacked spectrum does not show the presence of clear UV emission lines except for the \lya. The absence of these lines excludes the
possibility that obscured AGNs constitute a significant percentage of our LAEs.
Nonetheless, to further investigate the possibility of obscured AGNs, we match our catalogue with the one from \cite{Lyu+22} who extended the detection of AGNs in XDF by studying galaxies' rest-frame optical to mid-IR emission. Also in this case, we find no counterparts. 
Finally, we point out that LAEs hosting an AGN have a \lya\ luminosity $\rm L(Ly\alpha) \gtrsim 10^{43}~erg/s$ \cite[e.g.][]{Ouchi+08}. 
As displayed in Figure~\ref{fig:z-llya}, all our sources have a $\rm L(Ly\alpha) < 10^{43}~erg/s$. 
Based on these results, we safely exclude the presence of AGNs within our sample of 450 LAEs.

We then look for the UV/optical/near-IR counterparts of these objects.

\subsection{Photometric catalogue and SED fitting} \label{subsec:catalogue}
To find the UV/optical/IR counterparts of our sample of LAEs we first construct a source catalogue based on \hst\ and \jwst\ images, as follows.
We make use of the software \textsc{SExtractor} \citep{Bertin+96} to detect and measure photometry for sources in all the 28 available \hst\ and \jwst\ filters  that span the wavelength range $\lambda \simeq 0.2 - 5.6\mu\rm m$. 
We run \textsc{SExtractor} in dual-image mode, utilising a super-detection image that we create by combining photometric information from all the bands. 
To maximise the number of detected sources, we choose a \textit{hot-mode} extraction technique, as presented in \citet{Galametz+13}, which is well-suited to identify very faint sources.

We combine circular ($0.5$ arcsec diameter) and Kron apertures \citep{Kron+80} to extract the photometry (i.e., \texttt{MAG\_APER} and \texttt{MAG\_AUTO}, respectively). 
After applying aperture corrections $f_{\rm aper}$ to the \texttt{MAG\_APER} fluxes\footnote{The \hst\ aperture corrections for the different instruments and filters can be found at \url{https://www.stsci.edu/hst/instrumentation/}. 
For the \jwst\ filters, we estimate the aperture corrections using the \textsc{WebbPSF} software (\url{https://webbpsf.readthedocs.io/en/latest/}).} (see Table~\ref{tab:data_prop}), in each filter we select \texttt{MAG\_APER} over \texttt{MAG\_AUTO} for faint sources ($\geq 27$~mag) while, for brighter objects ($< 27$~mag), we adopt \texttt{MAG\_APER} (\texttt{MAG\_AUTO}) if \texttt{MAG\_APER} $<$ \texttt{MAG\_AUTO} (\texttt{MAG\_APER} $\geq$ \texttt{MAG\_AUTO}). 
Following \cite{Rinaldi+23a}, we set the magnitude threshold at 27~mag after performing various tests with the \hst\ photometry and comparing our fluxes with the HLF photometric catalogue. 

Due to the typical underestimation of photometric errors by \textsc{SExtractor} \citep[e.g.][]{Sonnett+13}, we set 0.05 mag as the minimum error for all the \hst\ photometry. 
We also set 0.05 mag as the minimum error for the \jwst\ photometry to account for possible uncertainties in the NIRCam and MIRI flux calibrations.

Finally, we correct all fluxes for Galactic extinction following \citet[][hereafter SF11]{Schlafly+11}. 
For each filter in our dataset, we present the multiplicative correction factors adopted to correct the galaxies flux for Galactic extinction $f_{\rm ext}$ in Table~\ref{tab:data_prop}.

We adopt the code \textsc{LePHARE} \citep{Ilbert+06} to perform the SED fitting and determine the properties of the detected sources.  
Our libraries are based on the stellar population synthesis models proposed by \citet[][hereafter BC03]{Bruzual+03}, constructed considering the Chabrier IMF \citep{Chabrier+03} with a cut-off mass of 100 M$_\odot$ ($0.1 - 100$ M$_\odot$). 
For the stellar models, we take into account two distinct metallicity values: solar (Z = 0.02 = Z$_{\odot}$) and subsolar (Z = 0.004 = 0.2Z$_{\odot}$).
As for the star formation history (SFH), we adopt two different kinds of models: an \textit{instantaneous burst} model (i.e. a single stellar population model) and exponentially declining SFHs (known as \textit{$\tau$-model}, ${\rm SFR}(t)\propto e^{-t/\tau}$). 
In this last case, we adopt values of $\tau$ (the so-called \textit{e-folding time}) equal to 0.001, 0.01, 0.03, 1, 2, 3, 5, 8, 10, 15 Gyr.
We also complement the \citetalias{Bruzual+03} stellar templates with the empirical QSO templates available in \lephare\ from \cite{Polletta+06}. 
To take into account the effects of internal dust extinction, we allow the code to convolve each synthetic spectrum with the attenuation law by \citet[][hereafter C00]{Calzetti+00} and with the extrapolation proposed by \cite{Leitherer+02} at short wavelengths, leaving the colour excess E(B-V) as a free parameter with values ranging between 0 -- 1.5 in steps of 0.1. 
We run \lephare\ in the redshift range $z = $~0 -- 20 in a mode that takes into account the possible presence of nebular emission lines.
In this last case, \lephare\ accounts for the contribution of emission lines such as \lya, H$\alpha$, H$\beta$, [OII]$\lambda3727$\AA\ whose contributions to the SED are derived via the SFR -- luminosity conversions by \cite{Kennicutt+98}. 
Also the [OIII]$\lambda\lambda4959,5007$\AA\ doublet is included in the above-listed transitions, considering different ratios with respect to the [OII] line \citep{Ilbert+06}.
Finally, emission lines are considered only for galaxies with dust-free colour bluer than $|NUV - r| \leq 4$ and their intensity is scaled according to the intrinsic UV luminosity of the galaxy.

For undetected sources in a given filter, after masking all nearby sources, we place 1000 random non-overlapping circular apertures (0.5 arcsec diameter) on the sky region around each source and within a maximum distance of 15 arcsec from the source centre. After applying a $3\sigma$ clipping, we use the background r.m.s. (1$\sigma$) to estimate their flux upper limit. 
For \lephare, we use a 3$\sigma$ upper limit for these filters. 
In all the cases where we have no photometric information (e.g. the MIRI/F560W and NIRCam coverage areas differ), we set the flux value to $-99$.

After running \lephare, we clean our output catalogue for Galactic stars by cross-matching it with the Gaia Data Release 3 catalogue \citep{Babusiaux+22} and by excluding all sources that display a high stellarity parameter from \textsc{SExtractor} (i.e., \texttt{CLASS STAR}~$ > 0.8$) and lie on the stellar locus of the (F435W - F125W) vs (F125W - F444W) diagram, e.g. \cite{Caputi+11}.

To assess the quality of our SED fit, we compare the photometric redshift $z_{\rm phot}$ derived with \lephare\ to the catalogue of spectroscopic redshifts $z_{\rm spec}$ from MUSE \citep{Bacon+23} and \textit{JADES} \citep{Bunker+23}. 
For the MUSE spectroscopic redshifts, we limit our comparison to all those sources having a $z_{\rm spec}$ quality flag QF~$= 3$, i.e. the highest confidence estimates. 
We adopt 0.2 arcsec as a matching radius between our catalogue and those from MUSE and \textit{JADES}.
We find that only about 15\% of all the matched sources are catastrophic outliers, i.e. $|\Delta z|/(1+z_{\rm spec}) \geq 0.15$ with $\Delta z = z_{\rm phot}-z_{\rm spec}$, and the normalised median absolute deviation $\sigma_{\rm NMAD} = 0.04$ ($\sigma_{\rm NMAD} = 1.48 \times {\rm median}\{|\Delta z - {\rm median}(\Delta z)|/(1+z_{\rm spec})\}$).

\subsection{HST/JWST counterparts for the Lyman-$\alpha$ emitters} \label{subsec:lya_counterparts}
To find the \textit{HST/JWST} counterparts of the LAEs in XDF, we match them with our photometric source catalogue (see Section~\ref{subsec:catalogue}) within a circular aperture of 0.5 arcsec radius.   
We find that 72 LAEs (about 16\% of the whole sample) do not have a counterpart in our catalogue, 250 (about 56\%) match with a single source and 128 (about 28\%) have multiple sources in their vicinity (up to five possible counterparts). 
Similar percentages were also reported by \cite{Bacon+23} who found that 15\% of their sample had no \hst\ counterpart, while 68\% were matched to a single object.
Upon examining the cutouts of LAEs lacking counterparts in our catalogue, we observe that these objects typically exhibit detections solely in the filter corresponding to the \lya\ emission, and in some cases, they show no counterparts in any filter. 
This pattern implies that these objects are likely low-mass, low-metallicity, high-equivalent-width emitters that are too faint to be detected even in deep imaging \cite[$m_{\rm AB} \gtrsim 30$~mag; e.g.][]{Maseda+18, Maseda+20, Mary+20, Maseda+23}. 
In scenarios where no detection is observed in any filter, an alternative explanation could be that the counterpart of these LAEs lies at a projected distance exceeding 0.5 arcsec from their MUSE detection. In such cases, the separation between the \lya\ emission and the UV/optical counterpart would correspond to an offset greater than 3.9 kpc (2.7 kpc) at $z = 2.8$ ($z = 6.7$). 
It is worth noting, however, that such offsets would be notably larger than what is typically observed at similar and lower redshifts, as indicated by previous studies \cite[e.g.][]{Hoag+19,Rasekh+22,Ribeiro+20}.
Furthermore, LAEs with offsets exceeding 3-4 kpc (larger than the typical size of galaxies at those redshifts) constitute a relatively small fraction of all LAEs and are typically associated with UV-bright systems \cite[e.g. ][]{Lemaux+21}.

To ensure accurate identification of the LAE counterpart, we re-run the photometric catalogue of all matched sources through \lephare, fixing this time the redshift of each counterpart candidate to that of the corresponding LAE. 
In fact, due to the crowding of the field and the depth of the observations, within the matching radius of 0.5 arcsec some foreground and background sources could be wrongly associated with the LAE.
Besides, the detection of one, or more, close-by sources (on the sky-plane) to the peak position of the \lya\ emission does not necessarily identify the LAE counterpart.
Because of this, we decide to only retain the sources with a $\chi^2_{\rm red}$ value for the best SED fit (after fixing the redshift to the spectroscopic one) below a given threshold. 
To determine this threshold, we adopt the value corresponding to the 84th percentile of the $\chi^2_{\rm red}$ distribution of LAEs which are matched uniquely to one single counterpart, i.e. $\chi^2_{\rm red} =  6.3$.  

Since we aim at deriving the physical properties of LAEs from the SED fitting, we limit our study sample to all sources with a secure detection in the MIRI filter ([F560W] $< 29.5$ mag).
The detection in the MIRI filter allows us to robustly constrain the optical/near-IR emission of our targets ($0.8 - 1.6~\mu$m, depending on redshift). 
This requirement ensures a more robust estimate of their stellar mass and the age of their underlying stellar population (for more details see Appendix~\ref{appdx:f560w_impact}).

Finally, we verify that different LAEs are not associated with the same counterpart.

Our final sample consists of 222 LAEs.
We show the position of these galaxies in Figure~\ref{fig:rgb_ima}.
Out of these final 222 objects, 182 LAEs (about 82\%) are associated with a single source, 34 (about 15\%) have two counterparts and 6 (about 3\%) have three counterparts.
We limit our statistical analysis only to the 182 sources with a single counterpart because of their complex interpretation. 
In fact, the multiple components are easily characterised by different physical properties (e.g. stellar mass, dust extinction, age of the stellar population) both in the case of clumps and different gravitationally bound/interacting systems. 
Besides, the association of such multiple components to the \lya\ emission is not trivial, especially when only based on photometric data.

\begin{figure}[t!]
    \centering
    \includegraphics[width=\columnwidth]{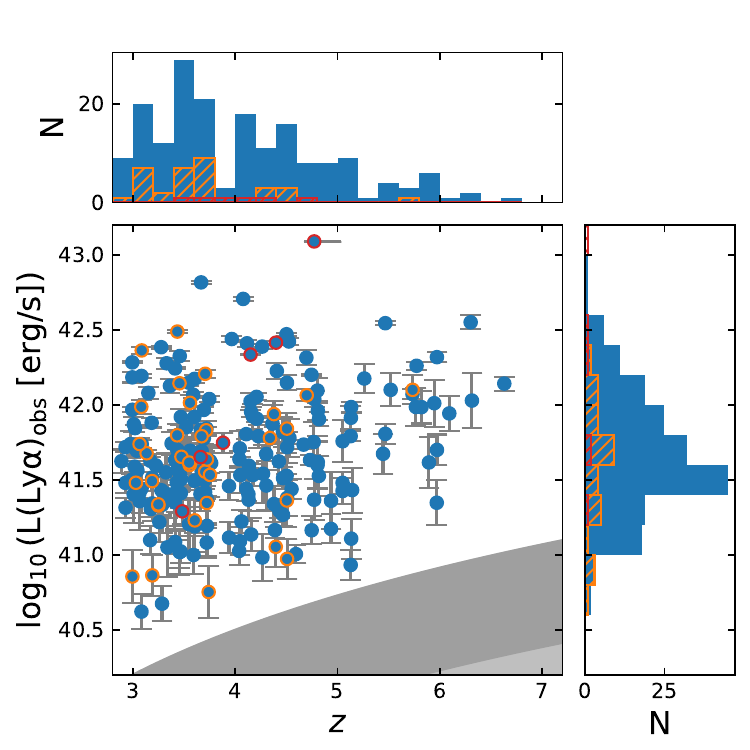}
    \caption{Observed \lya\ luminosity vs. redshift diagram of our final sample of LAEs. We highlight with orange and red edges LAEs matched to two and three photometric counterparts, respectively. 
    The top and right panels show the redshift and \lya\ luminosity distribution of our source. The grey shaded region indicates the luminosity corresponding to the 2$\sigma$ depth of the MUSE observations \citep{Bacon+23}, i.e. $2.1 \times 10^{-19}$ and $4.2 \times 10^{-20}$~erg/s/cm$^2$ at 10- and 141-h depths, respectively.}
    \label{fig:z-llya}
\end{figure}

For the LAEs matched to a single counterpart, the median offset $\delta_{\rm Ly\alpha}$ between the coordinates of the UV/optical counterpart and the peak of the \lya-emission is about 0.1 arcsec. Converting the offset separation for each source from arcsec into kpc, we derive a median value of about 0.8 kpc.
These estimates are broadly consistent with the typical \lya\ -- UV continuum offsets found in previous work targeting LAEs both at lower \cite[e.g.][]{Rasekh+22} and similar redshifts \cite[e.g.][]{Hoag+19, Ribeiro+20, Claeyssens+22, Iani+21, Iani+23b}. 
We present the redshift and observed \lya\ luminosity distribution of our final sample of LAEs in Figure~\ref{fig:z-llya} and \ref{fig:z-dist}.

\subsection{Lyman-Break Galaxies in XDF} \label{subsec:lbgs_selection}
In addition to LAEs at $z\simeq 3-7$ and based on our photometric catalogue, we also define a sample of sources in the same redshift range of our LAEs but that do not display any sign of \lya-emission (at least according to the available MUSE observations). 
According to recent literature \cite[e.g.][]{Dayal+12, deLaVieuville+20, ArrabalHaro+20}, in the following we refer to these objects as LBGs. 
The comparison of the physical properties of these sources with those of LAEs could determine if these two classes of objects are well distinguished or show common properties (see Section~\ref{subsec:laes_lbgs}).

\begin{figure}
    \centering
    \includegraphics[width = .9\columnwidth]{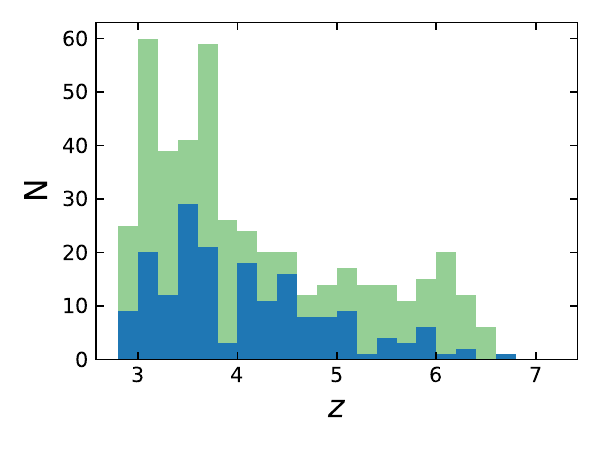}
    \caption{Redshift distribution of our sample of LAEs (in blue) and LBGs (in green).}
    \label{fig:z-dist}
\end{figure}

To this aim, we select all the sources in our photometric catalogue that have the best photometric solution from \lephare\ (\texttt{Z\_BEST}) within the $z$-range covered by our sample of LAEs.
We additionally discard all the sources with a best-fit $\chi^2_{\rm red} > 6.3$ (as for the LAEs' sample) and that have a clear preference ($\Delta\chi^2_{red} > 4$) for a stellar or an AGN model.
To further exclude AGNs, we additionally remove all matches ($\leq 1$ arcsec) between our LBGs catalogue and the AGN catalogues from \cite{Ranalli+13, Luo+17, Evans+20, Lyu+22}. 
We also discard all sources that are matched or close ($\leq 1$ arcsec) to a LAE reported in \cite{Bacon+23} (i.e. $\rm QF = 1, 2, 3$), and exclude those whose light is contaminated by a nearby source. 
Lastly, similarly to the LAEs sample, we limit our study to objects with detection in the MIRI/F560W filter and a magnitude $< 29.5$ mag.  
By doing so, we end up with a sample of 450 LBGs.
We present the redshift distribution of the selected LBGs in Figure~\ref{fig:z-dist}.

\begin{figure*}
    \centering
    \includegraphics[width = .7\textwidth]{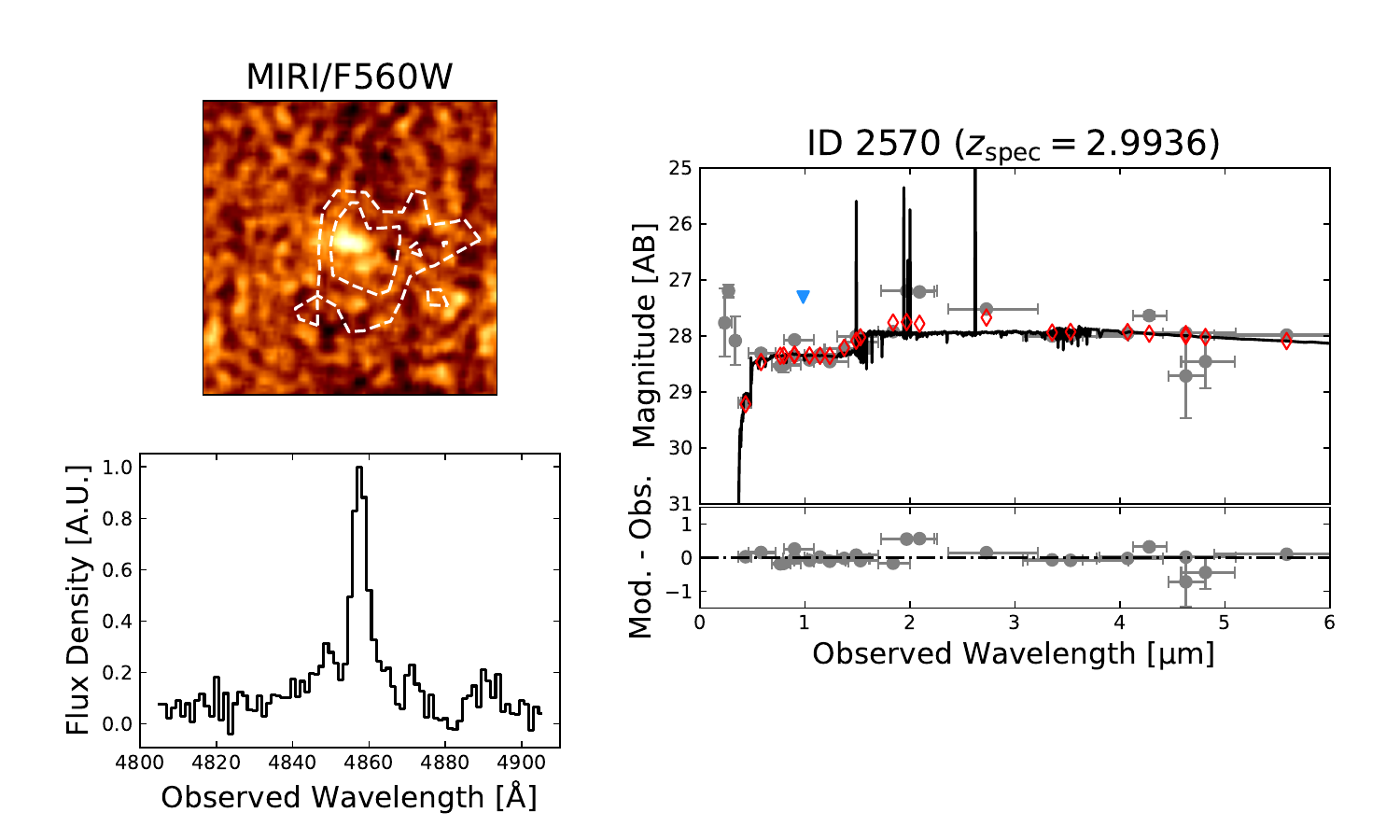}
    \includegraphics[width = .7\textwidth]{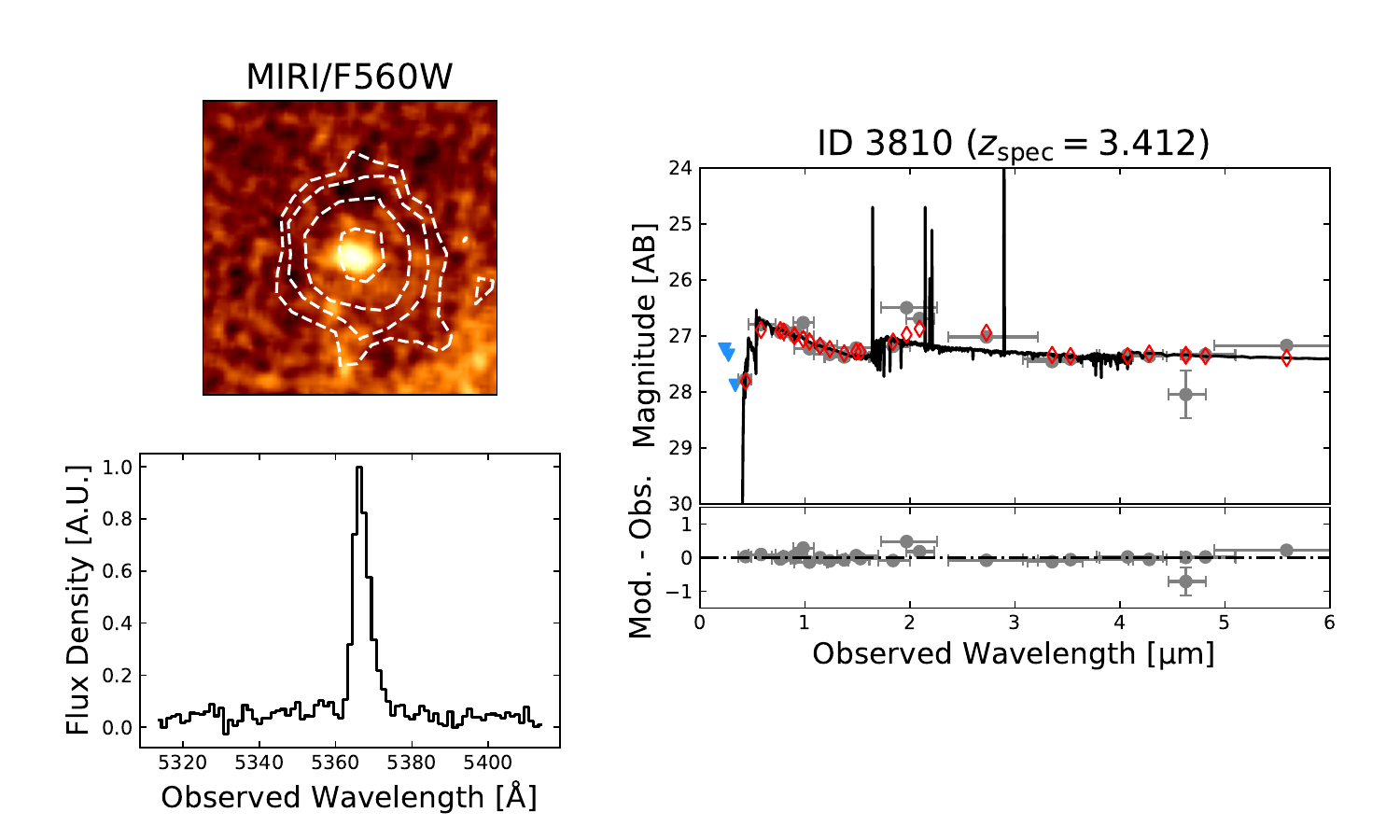}
    \includegraphics[width = .7\textwidth]{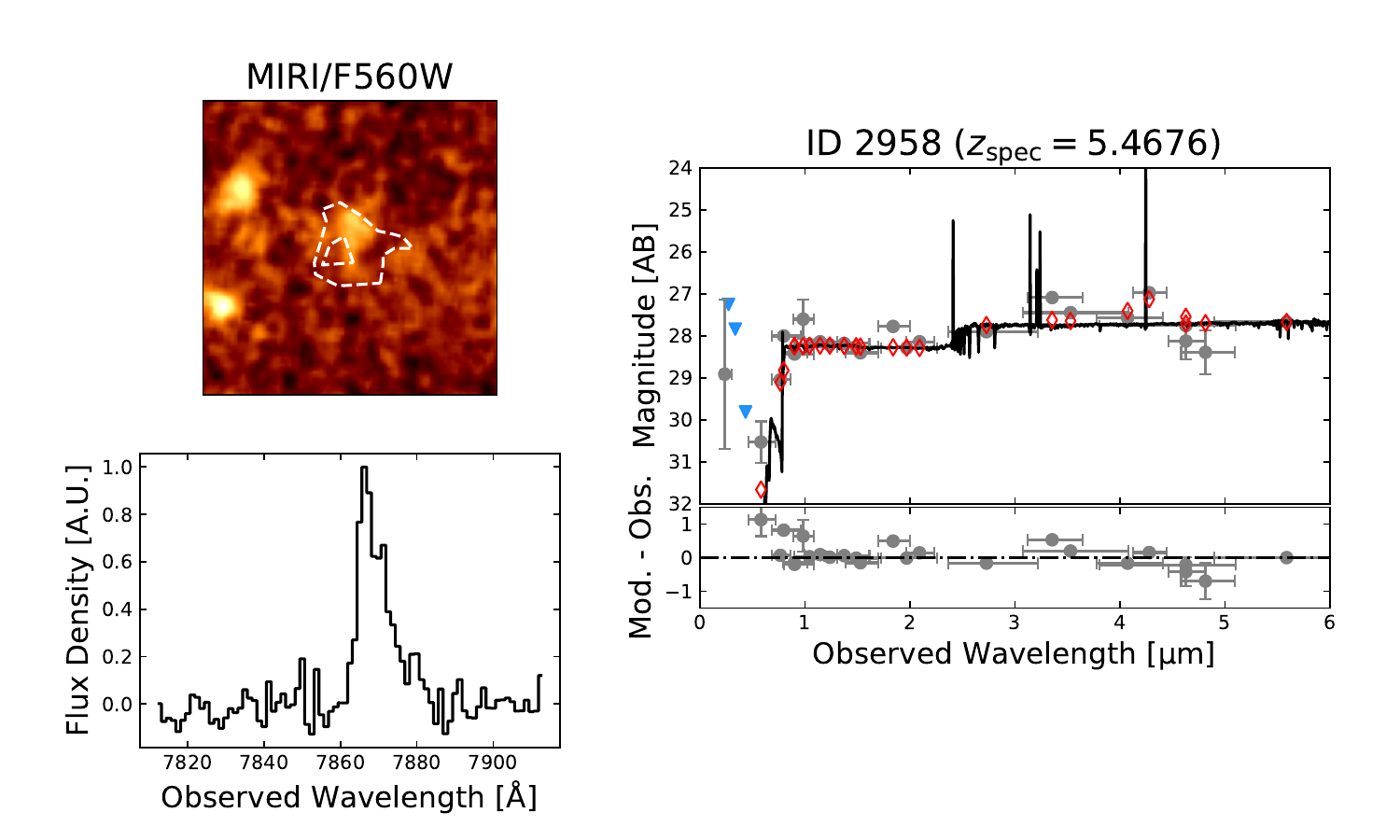}
    \caption{Examples of the imaging and photometric properties of three LAEs in our study: ID 2570 (top), ID 3810 (centre), ID 2958 (bottom). For each LAE, in the top left corner, we present  MIRI/F560W cutout images ($2.5'' \times 2.5''$) with overplotted \lya\ contours (in white) as derived from MUSE. In the bottom left panels, we show the integrated \lya\ spectrum of the galaxies while, in the right panels, we present the LAEs photometry (in grey with upper limits in blue) along with the \lephare\ best-fit SED (black) and synthetic photometry (red open diamonds).}
    \label{fig:laes_examples}
\end{figure*}

\section{Physical properties of the LAE counterparts} \label{sec:laes_prop}
In the following Section, we present and discuss the main physical properties of the LAEs based on the SED analysis of their UV/optical/near-IR counterparts.
In Figure~\ref{fig:laes_examples}, we show a few examples of the LAEs investigated in this work.

\subsection{SED properties of the LAEs}
\label{subsec:laes_prop_sed}

\begin{figure*}[t]
    \centering
    \includegraphics[width=.32\textwidth]{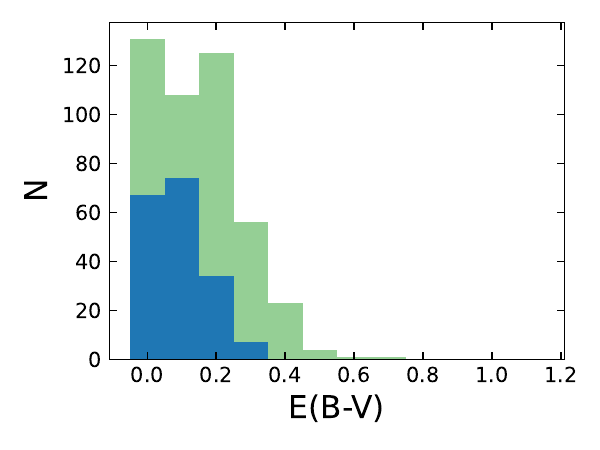}
    \includegraphics[width=.32\textwidth]{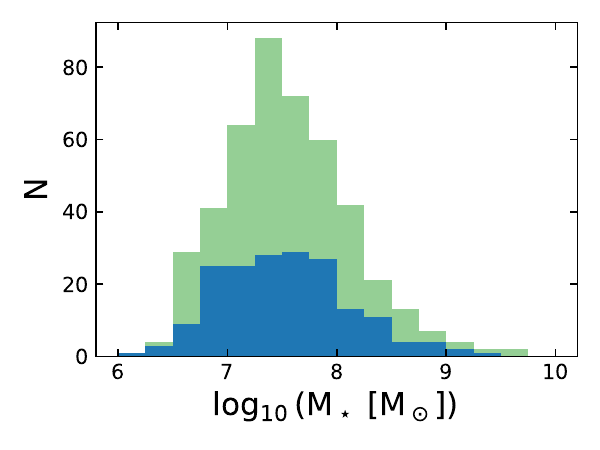}
    \includegraphics[width=.32\textwidth]{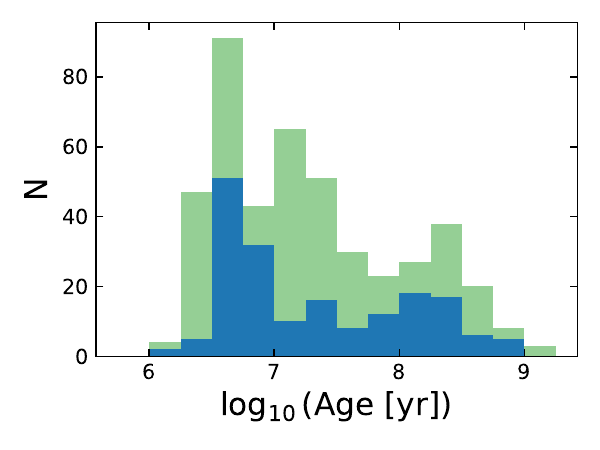}
    \includegraphics[width=.32\textwidth]{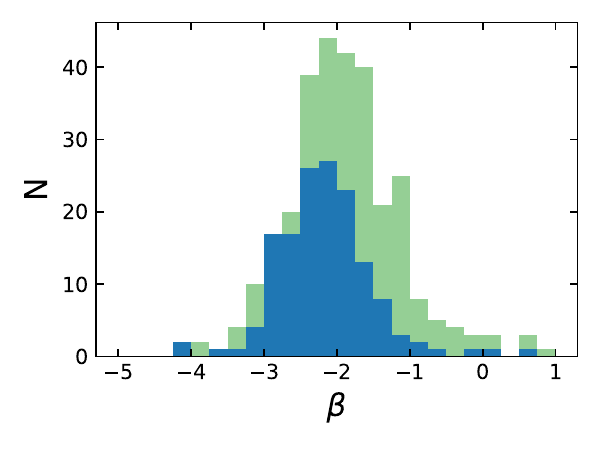}
    \includegraphics[width=.32\textwidth]{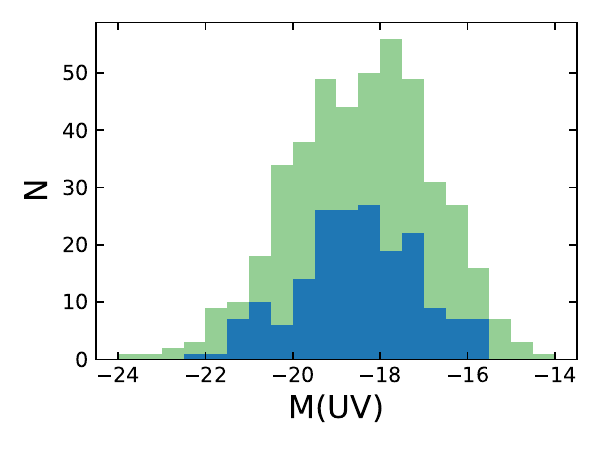}\\
    \includegraphics[width=.32\textwidth]{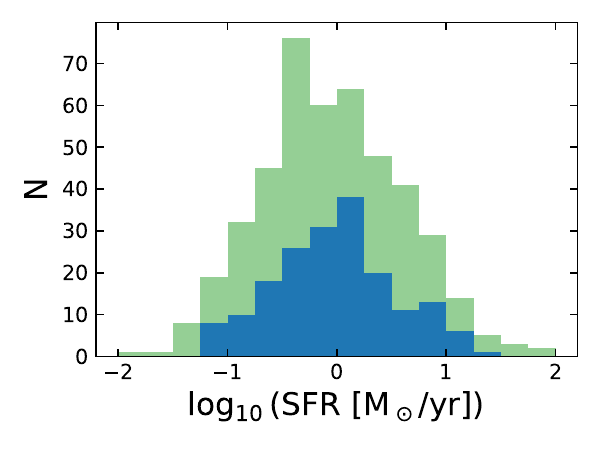}
    \includegraphics[width=.32\textwidth]{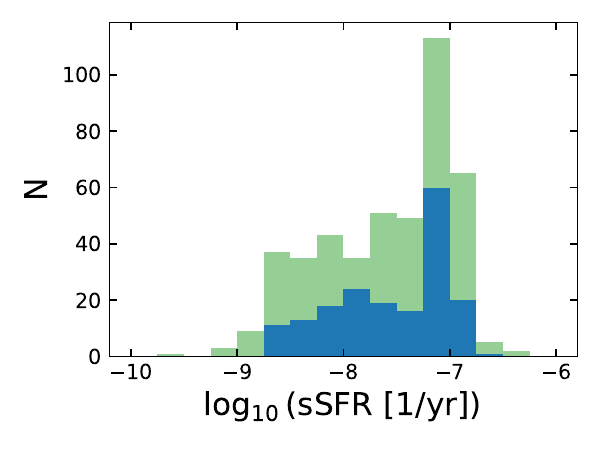}
    
    \caption{Distributions of the physical properties for our sample of LAEs (in blue) and LBGs (in green). In the top row we present the colour excess, stellar mass and age as derived from the \lephare\ SED fitting, while in the central and bottom rows, we display the distribution of the parameters directly derived from the analysis of the UV photometry of our sources, i.e. the UV continuum slope ($\beta$), the dust-corrected UV absolute magnitude M(UV), the star formation rate SFR and the specific star formation rate sSFR.}
    \label{fig:sed_prop_hist}
\end{figure*}

Having identified the right counterpart for each LAE (see Section~\ref{subsec:lya_counterparts}), we inspect the properties derived from the SED fitting: the colour excess \ebv, the stellar mass \mstar, and the age of the stellar population.
In Figure~\ref{fig:sed_prop_hist} we present the distributions of these quantities for our sample of LAEs.

We find that the bulk ($\simeq 77$\%) of our LAEs has \ebv~$\lesssim 0.1$ in agreement with previous studies \cite[e.g.][and references therein]{Ouchi+20}. 
Interestingly, however, the \ebv\ distribution presents a tail reaching values up to \ebv~$= 0.3$ that corresponds to a visual extinction $A_V \simeq 1.2$~mag, if we assume the \citetalias{Calzetti+00} attenuation law. 

\begin{figure}[t]
    \centering
    \includegraphics[width = .9\columnwidth]{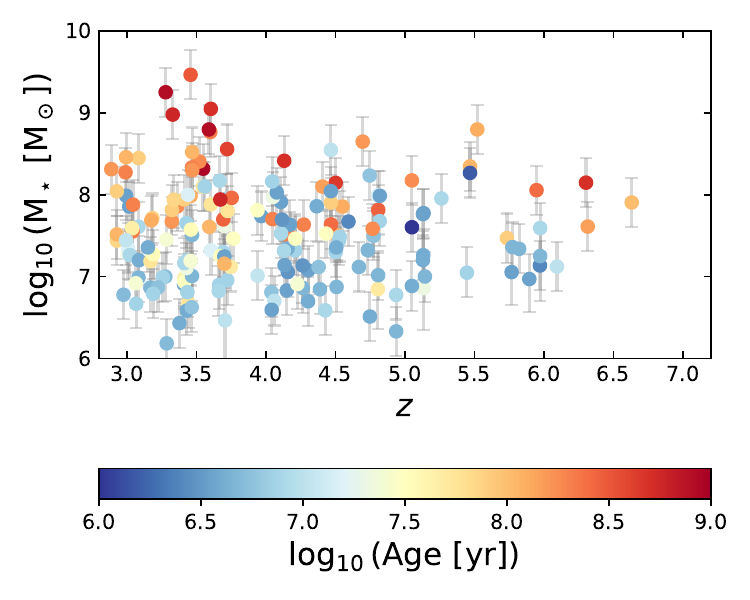}
    \caption{The stellar mass M$_\star$ distribution of our LAEs as a function of redshift $z$. The representative points of our sample of LAEs are colour-coded according to the best-fit age of their underlying stellar populations.}
    \label{fig:z_mass_age}
\end{figure}

As for the stellar mass, LAEs in our sample have masses spanning the range between a few $10^6$ to a few $10^{9}~{\rm M_\odot}$ with a median value of about $10^{7.7}~{\rm M_\odot}$. 
All in all, the LAEs in our sample are all low-mass systems.
In Figure~\ref{fig:z_mass_age} we show the LAEs stellar mass as a function of redshift \cite[][]{Fynbo+01, Nakajima+12, Hagen+14, Napolitano+23}. 
While the few most massive galaxies with $\rm M_\star \geq 10^{9} ~M_\odot$ in our sample are typically at $z < 4$, galaxies with a stellar mass between  $10^8 \rm~ M_\odot$ and  $10^9 \rm~ M_\odot$ are found up to $z \simeq 6$.

The age distribution of our objects is bimodal.
While 75\% of all the LAEs are best-fit by a young stellar population (age $< 100$~Myr), the remaining 25\% appear to have significantly older stellar populations, reaching in a few cases an age of 1 Gyr. This is in line with previous findings in the literature \citep[e.g.][]{Lai+08, Finkelstein+09, Pentericci+09, Rosani+20, Napolitano+23}.
Interestingly, we find LAEs with an underlying older stellar population already at $z \simeq 6.5$ (see Figure~\ref{fig:sed_prop_hist}), thus suggesting that 800 Myr after the Big Bang there were already galaxies characterised by an old stellar population and undergoing a new phase of star-formation \cite[\textit{rejuvenation}, e.g.][]{Rosani+20}.
We discuss the differences between the properties of LAEs with an underlying young and old stellar population further in Section~\ref{subsec:young_old_laes}.

We finally highlight that the majority (71\%) of our LAEs sample tends to prefer sub-solar stellar metallicities, while the remaining LAEs (29\%) have SEDs best-fitted with solar metallicity templates.
Although metallicities based on SED fitting can be highly uncertain, this result is broadly consistent with previous studies which found that LAEs are typically metal-poor systems \cite[][]{Finkelstein+11, Nakajima+12, Trainor+16, Kojima+17}.

As a final test, we inquire if the results reported above could be affected by the best fit SFH selected by \textsc{LePHARE} during the SED fitting. We, however, do not find any correlation between the SFH and the value of the different parameters retrieved. For more details see Appendix~\ref{appdx:impact_sfh}.

\subsection{UV continuum slope and absolute magnitude}
\label{subsec:beta_and_muv}
In addition to the above parameters, we investigate the observed slope of the UV continuum $\beta$ and the absolute UV magnitude M(UV) of our sample.
To estimate the $\beta$-slope we apply the methodology presented in \cite{Castellano+12}, i.e. fitting the observed magnitude covering the rest-frame UV emission of LAEs by means of the equation:
\begin{equation}
    m_{i} = -2.5\cdot(\beta + 2)\cdot \log_{10}(\lambda_{{\rm eff}, i}) + c
\end{equation}
where $m_{i}$ is the observed magnitude of the $i$th filter, $\lambda_{{\rm eff}, i}$ its corresponding effective wavelength, and $c$ is a constant representing the intercept for each best-fit observed UV continuum slope. 
For the estimate of $\beta$, we consider only broad band filters that cover the rest-frame wavelength range between 1300 and 2500 \AA. 
We reject all filters with a transmission curve covering rest-frame wavelengths below 1300~\AA\ to avoid any possible contamination from the \lya\ in the $\beta$-slope estimate.
We also discard the medium band filter \hst/F098M since it could be easily affected by the presence of other UV emission lines.
Thanks to the extensive dataset at our disposal, these conditions allow us to have between four to eight photometric bands (depending on redshift) to perform the  UV continuum fit of our sources.   
On a source-by-source basis, we further discard all filters with an upper limit and perform the fit only for those LAEs that retained at least three photometric bands with a detection \cite[e.g. ][]{Bolamperti+23}. 
These conditions are met for 148 sources (out of 182) of our sample (about 81\%).
For each one of these LAEs, we estimate $\beta$ and its associated error by drawing 1000 Monte Carlo realisation of their photometry, perturbing the observed magnitudes according to their errors and following a Gaussian distribution.
Then, we assume as $\beta$ the median of the final distribution of all the 1000 Monte Carlo realisations while we adopt the standard deviation of the distribution as its associated error.
In Figure~\ref{fig:sed_prop_hist} we present the distribution of the estimated $\beta$-slopes for all the 148 LAEs.
The derived distribution has a median value $\beta = -2.21 \pm 0.56$. 

We estimate then the LAEs' UV absolute magnitude M(UV) at 1500~\AA. 
For the 148 sources with an estimate of $\beta$, we derive M(UV) directly extrapolating the absolute magnitude at 1500~\AA\ from the best-fit of the UV continuum slope. 
For the 34 remaining sources, we first find the median value of the photometric bands with a detection within the rest-frame wavelength range 1300 -- 2500 \AA, and then, we apply the distance modulus. 

The derived value, however, does not correspond to the absolute magnitude value at 1500~\AA\ but, approximately, at the median of the respective effective wavelengths $\bar{\lambda}$ of the filters adopted for the estimate. 
In this case, a correction factor has to be applied to the so-derived absolute magnitudes to bring them to the corresponding value at 1500~\AA. 
This correction factor $\delta$M depends, however, on the slope of the UV continuum according to the relation:
\begin{equation}
    \delta {\rm M} = -2.5 \cdot (\beta+2) \cdot \log_{10}\left( \frac{1500}{\bar{\lambda}~ {\rm [\AA]}} \right)
\end{equation}
Since we do not have an estimate of $\beta$ for these sources, we adopt the median value $\beta$. 
By doing so, we derive a median magnitude correction of $\delta {\rm M} = -0.08 ^{+0.18}_{-0.21}$~mag. 

Finally, we correct the derived M(UV) for dust extinction assuming \citetalias{Calzetti+00} attenuation law and the E(B-V) obtained from \lephare.
We present the final distribution of UV absolute magnitudes in Figure~\ref{fig:sed_prop_hist}.

\subsection{Star formation rate}
\label{subsec:sfr}
Due to the lack of spectroscopic data targeting any of the Balmer lines of our LAEs (e.g. H$\alpha$, H$\beta$), we estimate their star formation rate (SFR) from the conversion of the luminosity of their UV stellar continuum.
We highlight, however, that the SFR derived from the H$\alpha$ (SFR(H$\alpha$)) and the one inferred from the UV luminosity (SFR(UV)) are known to describe the star formation of a galaxy on different timescales \cite[e.g.][]{Kennicutt+98, Leitherer+99, Kennicutt+12, Calzetti+13}: while SFR(H$\alpha$) is indicative of the galaxy \textit{instantaneous} star formation activity, i.e. its SFR over a timescale of about 10 Myr after the onset of a burst, the SFR(UV) depicts the SFR, assumed to be continuous and well-behaved, over at least 100 Myr.

Following \cite{Kennicutt+12}, we first convert the dust-corrected UV absolute magnitude at 1500~\AA\ into monochromatic luminosity, i.e. $L_{\nu}(1500{\rm\AA}) = 10^{-0.4\cdot({\rm M(UV)} - 51.6)}$, and then in SFR via the relation:
\begin{equation}
    {\rm SFR(UV)}~[{\rm M_\odot/yr}] =  C_{\rm UV} \cdot \frac{L_{\nu}(1500{\rm\AA})}{[{\rm erg/s/Hz}]}
\end{equation}
where $C_{\rm UV} = 8.82\times 10^{-29}~\rm M_\odot~ yr^{-1}~ erg^{-1}~ s~ Hz$.
We present the distribution of the SFR(UV) in Figure~\ref{fig:sed_prop_hist}. 

Based on the SFR(UV) derived via the above equation, we then compute the specific star formation rate sSFR ($\rm = SFR/M_\star$).
We present the distribution of sSFR in Figure~\ref{fig:sed_prop_hist}. 
The sSFR distribution clearly shows a bimodal distribution \cite[e.g.][]{Rinaldi+22}, with the majority of our sources having a sSFR$~> 10^{-7.6}\rm~ yr^{-1}$.

\begin{figure*}[t]
    \centering
    \includegraphics[width=\textwidth]{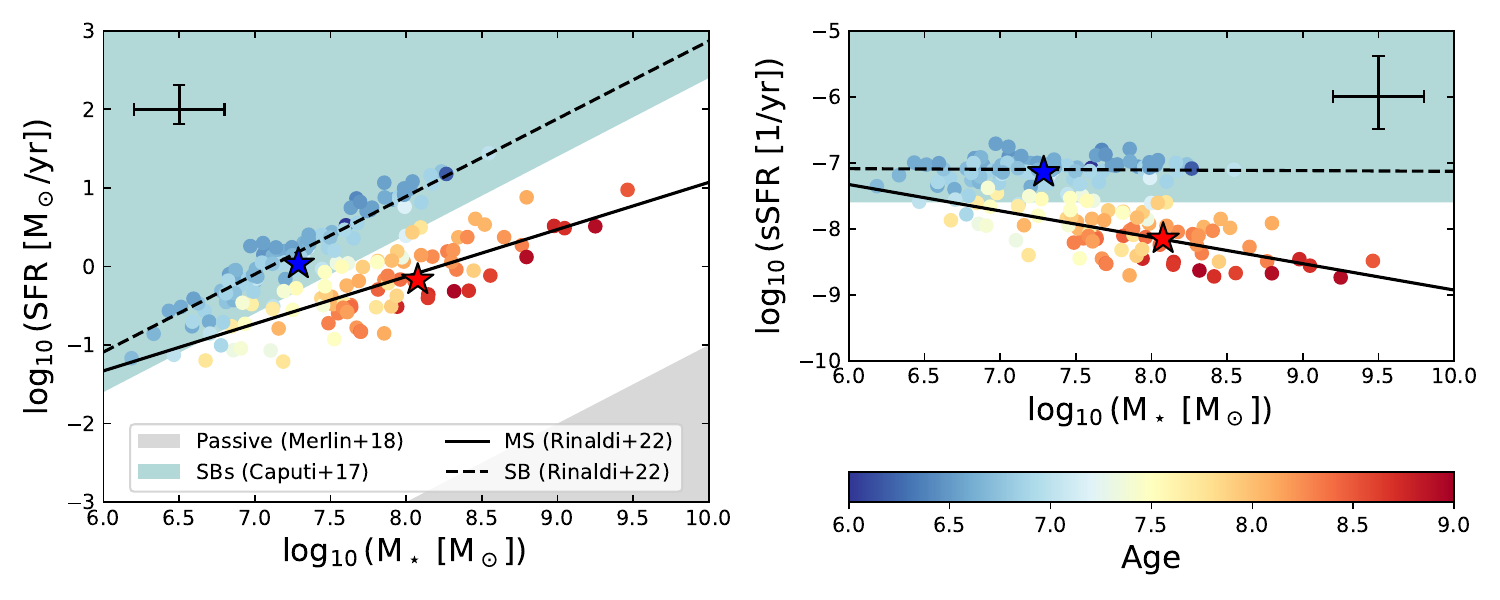}
    \caption{M$_\star$ -- SFR (left panel) and M$_\star$ -- sSFR (right panel) diagrams for our final sample of LAEs. The representative points of the LAEs are colour-coded according to the age of their stellar population. The blue and red stars represent the median value of M$_\star$ and SFR (sSFR) for LAEs showing an underlying young ($\leq 100$~Myr) and old ($> 100$~Myr) stellar populations, respectively. The blue shaded area is indicative of starburst (SB) galaxies \cite[${\rm sSFR} \geq 10^{-7.6}~\rm M_\odot/yr$, ][]{Caputi+17, Caputi+21} while the grey area highlights the locus of passive galaxies \cite[${\rm sSFR} \leq 10^{-11}~ {\rm M_\odot/yr}$, e.g.][]{Merlin+18}. The black solid (dashed) line shows the MS of star-forming galaxies (SB sequence) at $z \simeq 4$ as reported in \cite{Rinaldi+22}. The black errorbar in the top left corner is representative of the median errors of the LAEs sample.}
    \label{fig:ms_sfr}
\end{figure*}

Based on these results, we investigate which region of the M$_\star$ -- SFR plane our sample of LAEs populates. 
According to Figure~\ref{fig:ms_sfr}, we find that our LAEs display a bimodal distribution.
While the representative points of the oldest LAEs ($\gtrsim 100$~Myr) lie along the so-called Main-Sequence (MS) of star-forming galaxies at their redshift \citep{Rinaldi+22}, the youngest objects (the majority of our sample) are located above the MS, above the lower-boundary of the starburst galaxies \cite[SBs, ][]{Caputi+17, Caputi+21}.
The presence, at a given M$_\star$ value, of LAEs on the MS and in the SBs region as well as their separation in age, could be a sign that these objects went through different evolutionary paths, i.e. diverse star formation histories (SFHs), or the consequence of \textit{burstiness}, i.e. when the star formation is non-steady and out of equilibrium \cite[e.g.][]{Guo+16, Faisst+19, Atek+22}.

\begin{deluxetable*}{l|ccccccc}[t]
\tablenum{2}
\tablecaption{Table of the median values of the properties of LAEs with an underlying young ($<100$ Myr) and old ($\geq 100$ Myr) stellar population.}
\tablewidth{0pt}
\tablehead{
\colhead{} & \colhead{L(\lya)}  & \colhead{E(B-V)}  & \colhead{M$_\star$}  & \colhead{$\beta$} & \colhead{M(UV)} & \colhead{SFR(UV)} & \colhead{sSFR(UV)}\\
\colhead{} & \colhead{[erg/s]} &\colhead{}  & \colhead{$\log_{10}$([M$_\odot$])}  & \colhead{} & \colhead{[mag]} & \colhead{$\log_{10}$([M$_\odot$/yr])} & \colhead{$\log_{10}$([1/yr])}}
\startdata
young LAEs & $42.14_{-0.55}^{+0.69}$ & $0.11_{-0.11}^{+0.09}$ & $7.31_{-0.45}^{+0.58}$ & $-2.18_{-0.51}^{+0.59}$ & $-18.57_{-1.50}^{+1.33}$ & $-0.01_{-0.53}^{+0.60}$ & $-7.20_{-0.55}^{+0.16}$\\
old LAEs & $41.87 \pm 0.44$ & $0_{-0.0}^{+0.1}$ & $8.09_{-0.47}^{+0.54}$ & $-2.29_{-0.50}^{+0.55}$ & $-18.21_{-1.26}^{+1.09}$ & $-0.16_{-0.44}^{+0.51}$ & $-8.24_{-0.33}^{+0.26}$ \\
\enddata
\tablecomments{In the above table, we report in logarithmic value the \lya\ luminosity L(\lya), the stellar mass M$_\star$, the star formation rate SFR(UV) and the specific star formation rate sSFR(UV) and their corresponding 68\% scatter. We also present the values of E(B-V), $\beta$ and dust-corrected M(UV) with their 68\% uncertainties.}
\label{tab:young_old_median_prop}
\end{deluxetable*}

\begin{figure*}[t]
    \centering
    \includegraphics[width=.32\textwidth]{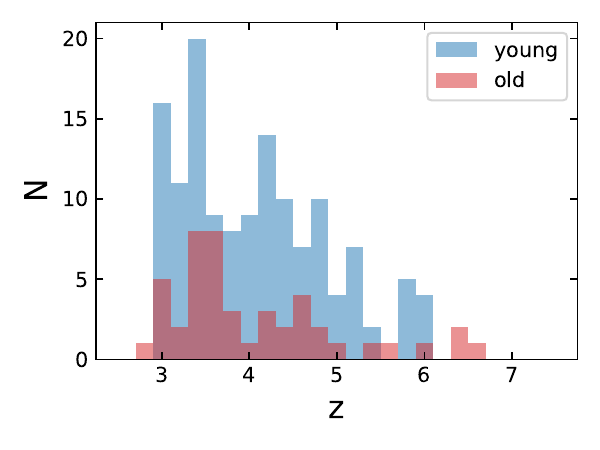}
    \includegraphics[width=.32\textwidth]{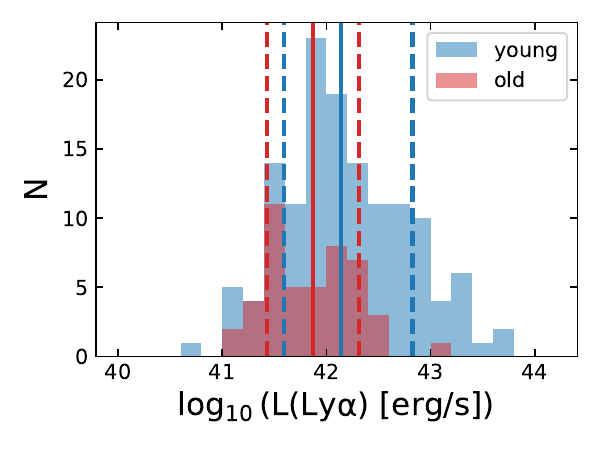}
    \includegraphics[width=.32\textwidth]{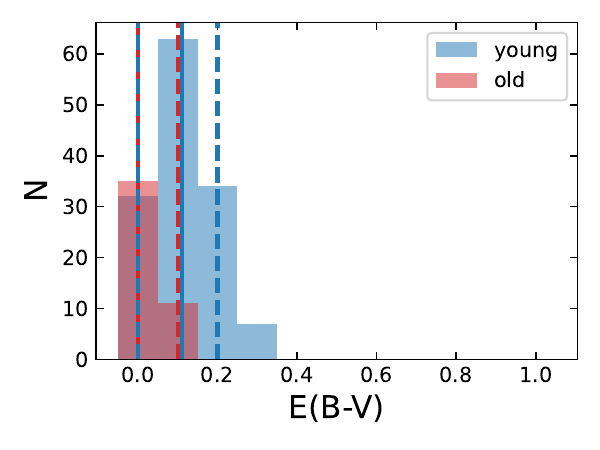}
    
    \includegraphics[width=.32\textwidth]{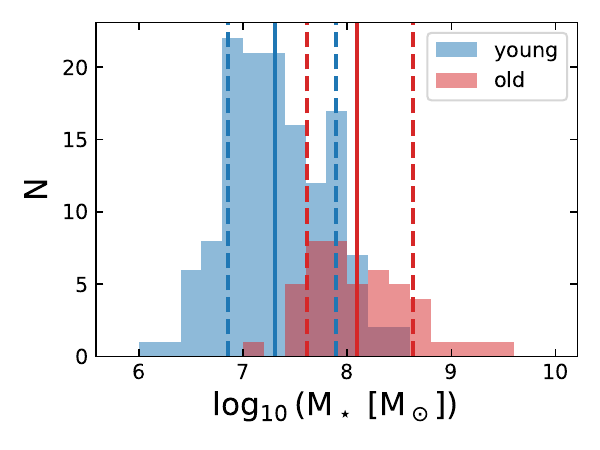}
    \includegraphics[width=.32\textwidth]{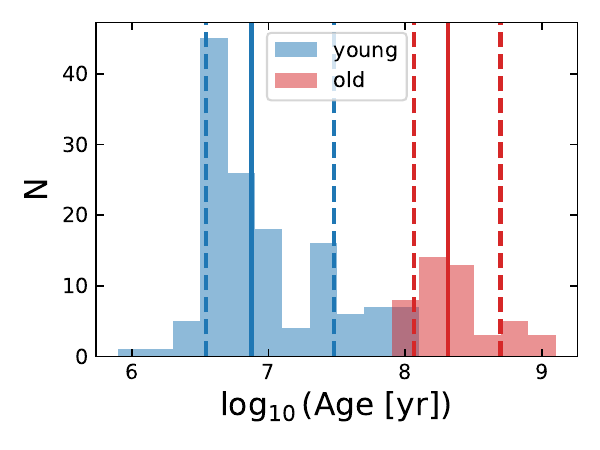}
    \includegraphics[width=.32\textwidth]{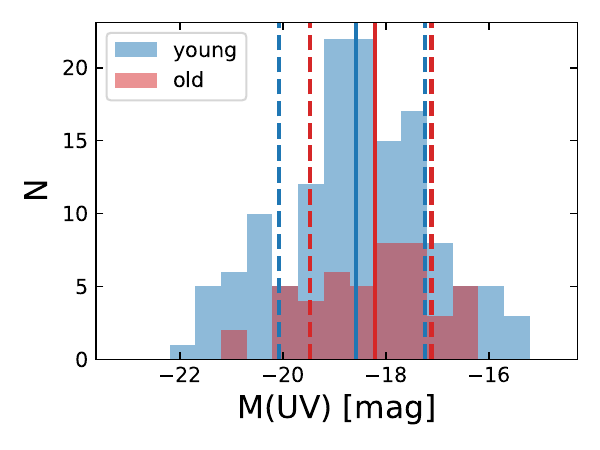}
    \includegraphics[width=.32\textwidth]{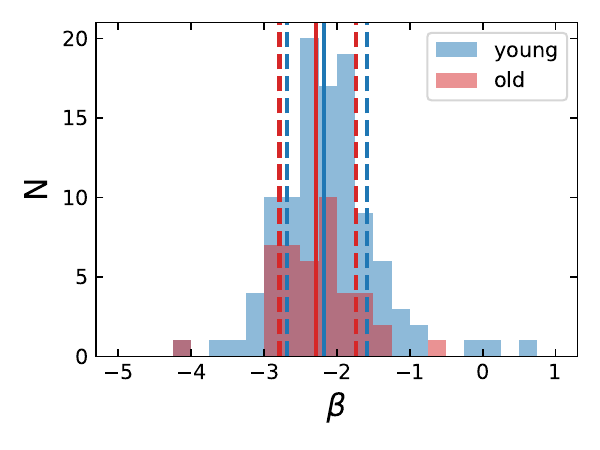}
    \includegraphics[width=.32\textwidth]{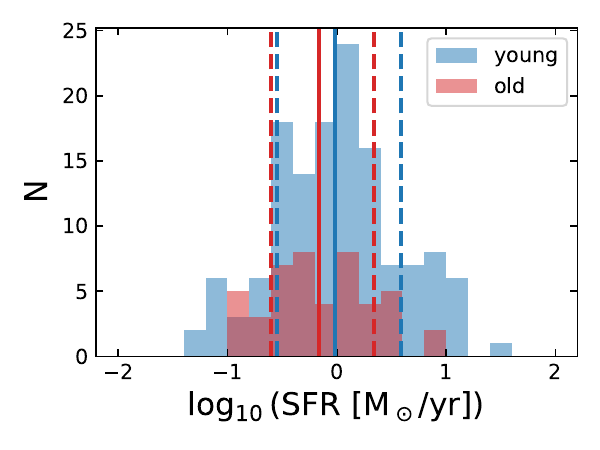}
    \includegraphics[width=.32\textwidth]{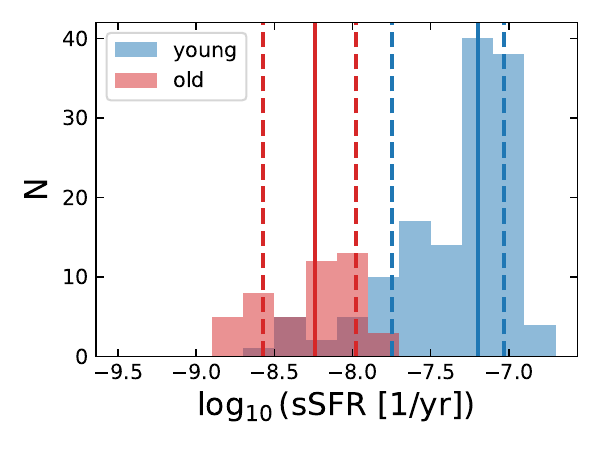}
    \caption{Distributions of the physical properties derived via the SED fitting for our final sample of LAEs separated into LAEs showing an underlying young ($< 100$~Myr, in blue) and old ($\geq 100$~Myr, in red) stellar populations. The vertical solid lines highlight the median value of the distributions, while the dashed lines are indicative of the 68\% confidence interval, i.e. the 16th and 84th percentiles, respectively.}
    \label{fig:old_new_laes_prop}
\end{figure*}

\begin{figure}
    \centering
    \includegraphics[width=\columnwidth]{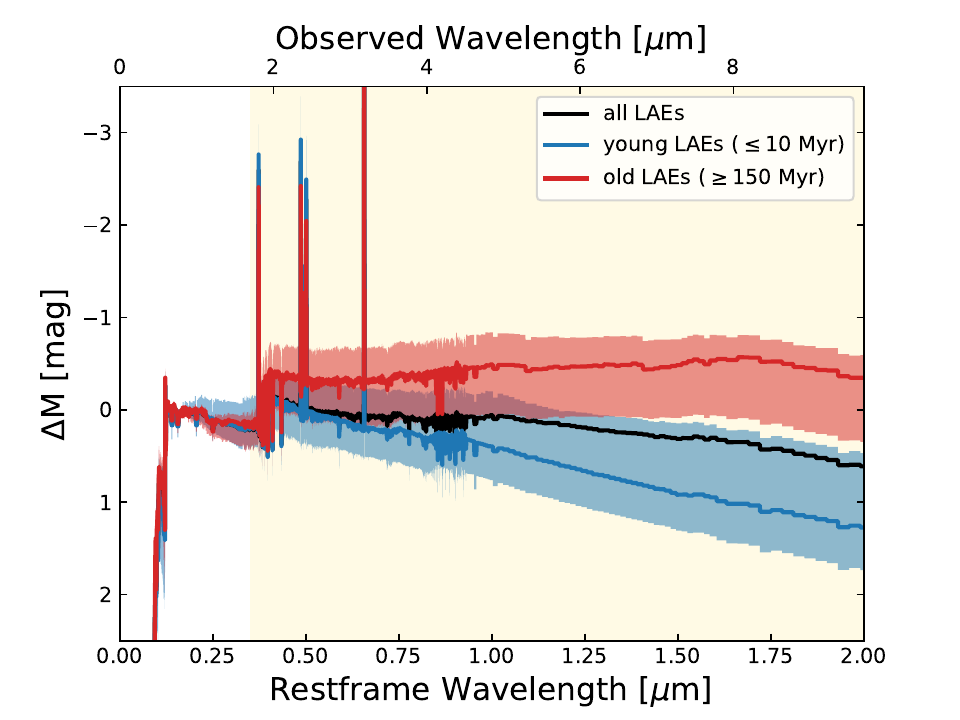}
    \caption{Average rest-frame SED (normalised at 1500 \AA) of the \lephare\ best-fits for young (age $\leq 10$~Myr, in blue) and old ($\geq 150$~Myr, in red) LAEs. The corresponding coloured shaded areas represent the 68\% confidence interval of the average SEDs, while the background yellow region is indicative of the wavelength range covered by the \jwst\ filters at $\lambda > 1.6~\mu$m assuming $z \simeq 4$.}
    \label{fig:sed_young-old}
\end{figure}

\begin{deluxetable*}{l|ccccccc}[t]
\tablenum{3}
\tablecaption{Two-sample KS test between the properties of LAEs with an underlying young ($< 100$~Myr) and old ($\geq 100$~Myr) stellar population.}
\tablewidth{0pt}
\tablehead{
\colhead{KS test} & \colhead{L(\lya)} & \colhead{E(B-V)}  & \colhead{M$_\star$} & \colhead{$\beta$} & \colhead{M(UV)} & \colhead{SFR(UV)} & \colhead{sSFR(UV)}}
\startdata
$p$-value & $0.04_{-0.03}^{+0.07}$ & 1e-8 & 1e-9 & $0.34_{-0.19}^{+0.24}$ & $0.43_{-0.25}^{+0.27}$ & $0.44_{-0.24}^{+0.28}$ & 1e-11\\
\enddata
\tablecomments{We report the exact estimate of the $p$-value only for parameters with $p$-value $\geq 0.05$ and for those values with an errorbar that comprises $p$-value $= 0.05$. In all the other cases, we only report its magnitude.}
\label{tab:ks_test_old_new}
\end{deluxetable*}

\subsection{Comparison between young and old LAEs}
\label{subsec:young_old_laes}

From the SED analysis (see Section~\ref{subsec:laes_prop_sed}) we found that the distribution of the LAEs' age is bimodal with 28\% of the overall sample having an underlying stellar population older than 100 Myr. 
This result hints towards the possible existence of two populations of LAEs with different properties \cite[e.g.][]{Shimizu+10, ArrabalHaro+20}. 
For this reason, we investigate if LAEs with an older stellar population display overall different general properties.

In Figure~\ref{fig:old_new_laes_prop} we present the distribution of the physical properties investigated in this paper separating LAEs with an underlying young ($\leq 100$~Myr, young LAEs hereafter) and old ($> 100$~Myr, old LAEs hereafter) stellar populations.
The histograms clearly show that old LAEs tend to be more massive, and less bright both in their \lya\ and UV luminosity if compared to young LAEs. 
The fainter absolute UV magnitude of young LAEs in turn converts into a systematically lower SFR (especially at fixed stellar mass), and a significantly lower sSFR ($\simeq 1$ dex), see Section~\ref{subsec:sfr}. 
We report the median values of the properties of LAEs with a young and old stellar population 
in Table~\ref{tab:young_old_median_prop}.

Our results are in line with those of \cite{ArrabalHaro+20}, who studied 404 LAEs at $z = 3-7$ with a clear rest-frame UV/optical detection and found that about 67\% of their sample was constituted by very young galaxies (median age $\simeq 30$ Myr) with stellar masses between $10^{8}-10^{9.5}~\rm M_\odot$, while the remaining 33\% was showing an overall older stellar population ($\simeq 1$ Gyr) and masses above $10^9~\rm M_\odot$.

For a more quantitative estimate of the differences between the two distributions of LAEs, we apply a two-sample Kolmogorov-Smirnov (KS) test.
For each physical quantity, we perturb 200 times the distribution of its values according to the errors. 
Then we run the two-sample KS test for all possible permutations of the randomly generated distributions, i.e. about 40000 realisations.
We assume as the final $p$-value the median of all measurements and consider its 68\% confidence interval.
We assume as a threshold a $p$-value~$ = 0.05$ to discern if the distributions descend from the same parent distribution (i.e. the \textit{null hypothesis}, $p \geq 0.05$) or not ($p < 0.05$).
For the investigated properties, we find that the two-sample KS test return values $p << 0.01$ for the colour excess, stellar mass and specific star formation rate, thus implying that young and old LAEs have indeed distinct distributions for these parameters. On the contrary, we find that the two classes of objects have similar distributions in UV continuum slope and absolute magnitude, star formation rate, and, marginally, \lya\ luminosity.
We present the results of the two-sample KS test in Table~\ref{tab:ks_test_old_new}.

The difference between the physical properties of the two classes of LAEs suggests a difference in their SED shapes.
In Figure~\ref{fig:sed_young-old} we present the rest-frame average best-fit SED of all the sources with a very young ($\leq 10$~Myr) and old ($> 150$~Myr) stellar population.
To obtain the average best-fit SEDs, we first bring every observed best-fit SEDs to the rest-frame.
Then, we resample them to a common wavelength range using the \textsc{Python} library \textsc{SpectRes} \citep{Carnall+17} and normalise them at 1500 \AA.
Finally, we derive the median trend and its corresponding 68\% confidence interval.

From a visual inspection of the so-derived median SEDs, the separation between the two classes of objects is clear when considering the \jwst\ dataset, i.e. photometry at $\lambda > 1.6 \mu$m.
The brighter rest-frame optical/near-IR fluxes, as well as the stronger Balmer Break (falling for these objects in the bluest \jwst\ bands), determine the higher mass and older age of old LAEs.  
On the contrary, when looking only at the wavelength regime covered by \hst, the photometric separation is small and well within the confidence interval of both median SEDs.
This result shows the fundamental role played by the \jwst\ near/mid-IR photometry in separating these two classes of objects. 
It is not possible to separate these two classes of LAEs based only on the available \hst\ information.

Finally, in line with previous studies \cite[e.g.][]{Nilsson+09, Shimizu+10}, we find that the percentage of old LAEs concerning the overall LAE population $\rm N_{old, LAE}/N_{tot, LAE}$ decreases with redshift. 
In fact, we find that $\rm N_{old, LAE}/N_{tot, LAE}$ decreases from $29_{-3}^{+1}$\% at $z=2.8-4$, to $20_{-4}^{+1}$\% at $z=4-5$, to $17_{-4}^{+5}$\% both at $z = 5-6$ and, finally, increases to $75_{-25}^{+1}$\% at $z=6-7$. 
The decreasing trend we retrieve for the first three redshift bins ($z = 2.8 - 6$) can be explained by considering that at higher redshifts, i.e. at younger cosmic ages, we expect to find less and less old galaxies. 
This diminishes the possibility of detecting galaxies going through a rejuvenation process at those redshifts.
As for the reversal trend at $z > 6$, we interpret it as a selection effect: young LAEs may be too faint to be detected even in the deep \jwst\ images available.
This scenario is clearly shown by the median SED of young LAEs presented in Figure~\ref{fig:sed_young-old} where the SED of young LAEs can have more than one magnitude difference at optical/near-IR wavelengths compared to old LAEs.

\section{Comparison between LAEs and LBGs} \label{subsec:laes_lbgs}
In the following section, we compare the results for our sample of LAEs to the more general population of galaxies at $z\simeq 3-7$ that do not display \lya\ emission. 
By doing so, we want to investigate if there is any clear separation between these two populations of galaxies based on the physical properties that can be inferred from their photometry.

We have described our LBG sample selection in Section~\ref{subsec:lbgs_selection}.
To derive the LBG physical properties we adopt the same methodology applied for LAEs, see Section~\ref{sec:laes_prop}.
From the SED fitting, we derive the LBG distribution in stellar mass,  age and colour excess.
We then derive their UV continuum slope, their absolute UV magnitude, their star formation rate and specific star formation rate. 

\begin{figure*}[t]
    \centering
    \includegraphics[width=\textwidth]{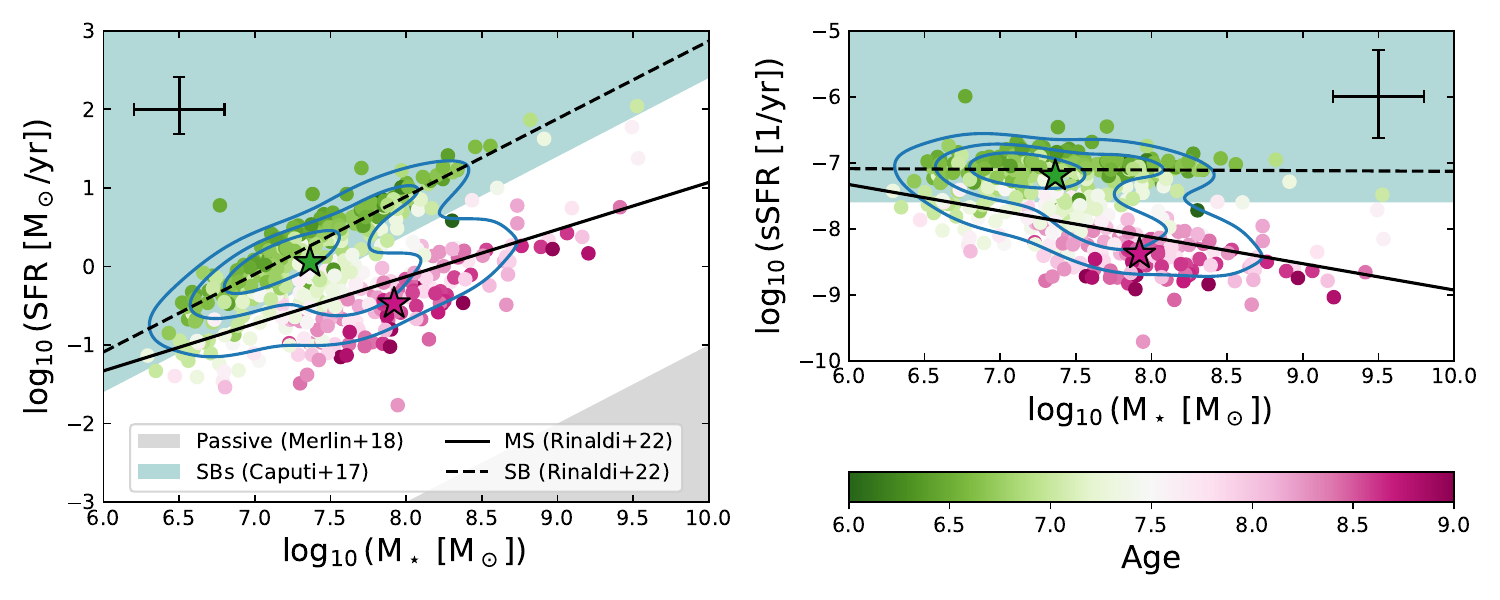}
    \caption{M$_\star$ - SFR (left panel) and M$_\star$ - sSFR (right panel) diagrams for our final sample of LBGs. The panels follow the same convention as in Figure~\ref{fig:ms_sfr}. 
    The superimposed blue contours are indicative of the area of the panels populated by our sample of LAEs.}
    \label{fig:ms_sfr_lbgs}
\end{figure*}

As a first step, we investigate the LBGs M$_\star$ -- SFR and M$_\star$ -- sSFR diagrams in Figure~\ref{fig:ms_sfr_lbgs}.
Similarly to LAEs, the LBGs show a clear bi-modal distribution in both diagrams, with the youngest objects mostly confined in the region of SB galaxies \citep{Caputi+17, Caputi+21} while the galaxies with an older underlying stellar population tend to crowd along the MS \citep{Rinaldi+22}.
LBGs and LAEs are also populating virtually the same region of the diagrams (see blue contours).
This suggests that their star formation histories are possibly similar and the galaxies assembled following analogous paths.

\begin{figure*}
    \centering
    \includegraphics[width = .32\textwidth]{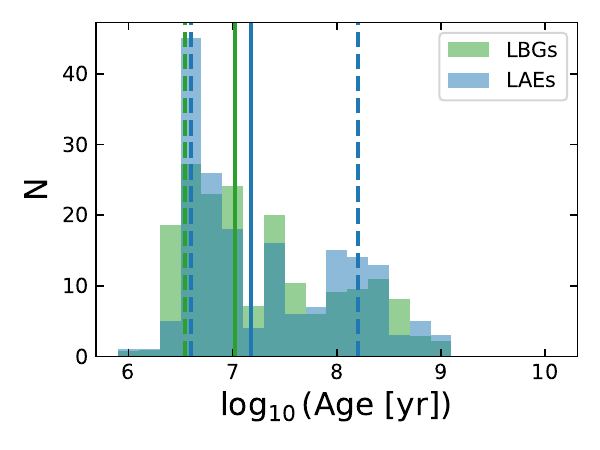}
    \includegraphics[width = .32\textwidth]{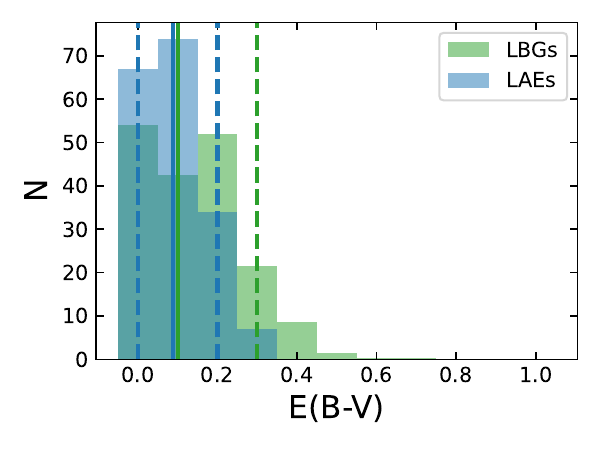}
    \includegraphics[width = .32\textwidth]{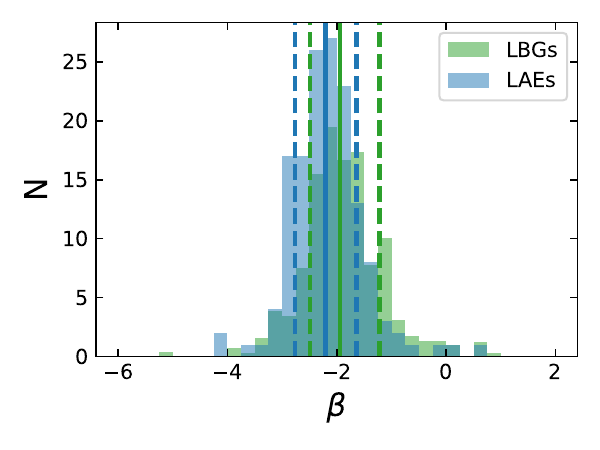}
    \includegraphics[width = .32\textwidth]{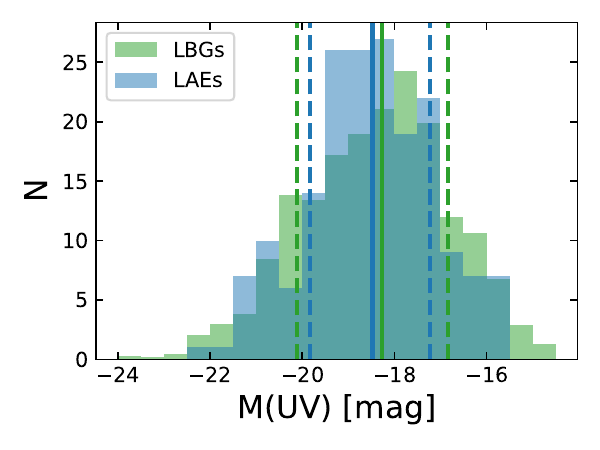}
    \includegraphics[width = .32\textwidth]{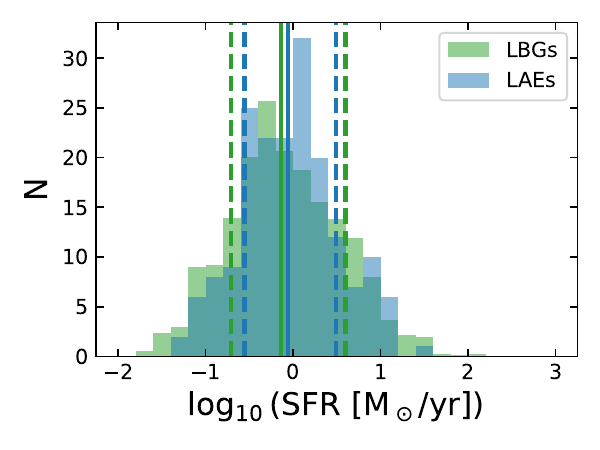}
    \includegraphics[width = .32\textwidth]{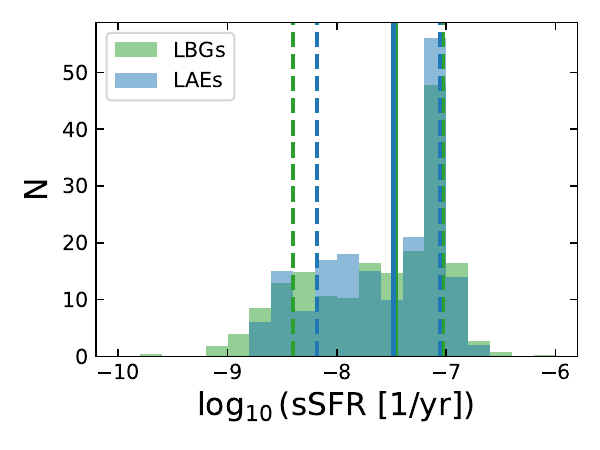}
    \caption{Comparison between the distribution of the physical properties of LAEs (in blue) and mass-matched LBGs (in green). The vertical solid lines highlight the median value of the distributions, while the dashed lines are indicative of the 68\% confidence interval, i.e. the 16th and 84th percentiles, respectively.}
    \label{fig:mass_matched}
\end{figure*}

Despite the fact that LBGs and LAEs cover almost the same stellar mass range (see Figure~\ref{fig:sed_prop_hist}), we mass-match the two samples to ensure an unbiased comparison of their physical properties. This means that, in the following, all the statistics that we present for the properties of LBGs will be weighted such that their stellar-mass distribution follows that of the LAEs.
We present the mass-matched distributions of the LBGs physical properties in Figure~\ref{fig:mass_matched}. 
For each measured parameter, we also report the median value and the 68\% confidence interval in Table~\ref{tab:laes_lbgs_prop}.

From a visual inspection of the panels in Figure~\ref{fig:mass_matched} and according to the values reported in  Table~\ref{tab:laes_lbgs_prop}, LAEs and LBGs appear to follow similar distributions for their properties: the median values of the different parameters for both classes of objects are comparable and fully within the 68\% confidence intervals of the distributions (see Table~\ref{tab:laes_lbgs_prop}).
For a more quantitative estimate, we run a two-sample KS test with a similar configuration to the one presented in Section~\ref{subsec:young_old_laes} and with only the addition of weights due to the mass-matching of the LBGs sample to the LAEs one.
We report the $p$-value and the corresponding error for each physical quantity in Table~\ref{tab:ks_test_laes_lbgs}.
Also in this case, we assume as a threshold value $p = 0.05$ for the \textit{null hypothesis}.
According to the two-sample KS test, LAEs and LBGs have very similar properties except for their distributions in E(B-V) and $\beta$ slope.
We note that a discrepancy between the E(B-V) distributions easily translates into a discrepancy between the $\beta$ slopes since the UV continuum is strongly and differentially affected by dust extinction.

The  stellar age distribution of LBGs (see top left panel of Figure~\ref{fig:mass_matched}) clearly shows that about 48\% of the LBG sample is constituted by galaxies with a young stellar population ($\leq 10$ Myr).  
Interestingly, this suggests that even though these systems are young they do not display the presence of the \lya\ emission. 
In Section~\ref{subsec:lya_lack}, we investigate whether this is simply due to the high dust extinction or whether HI resonant scattering could be playing a role in the absence of a \lya\ emission line.

Finally, we highlight that 55\% of the whole LBG sample is best-fit with a sub-solar template. We find a similar percentage ($\approx 49$ \%) even when considering only the population of young LBGs ($< 10$ Myr). 
These results are broadly comparable to what was found for LAEs for which the percentage of best-fit sub-solar templates is $\approx$ 71\% for both the overall sample and among the youngest objects.

\subsection{Understanding the lack of \lya\ emission in young LBGs}
\label{subsec:lya_lack}

The analysis of our sample of LBGs shows that 48\% of these galaxies are characterised by a young stellar population ($\leq 10$~Myr).
At the same time, however, they do not display \lya\ emission. 
To understand what could be the reason behind the lack of detection of their \lya, we attempt to estimate their expected observed \lya\ flux $\rm F(Ly\alpha)_{obs}$ and compare it with the depth of the MUSE observations available.

On the basis of the SFR -- L(\lya) conversion by \cite{Sobral+19}, we estimate $\rm F(Ly\alpha)_{obs}$ via the equation:
\begin{equation} \label{eq:lya_lbgs}
    \rm F(Ly\alpha)_{obs} = \frac{8.7 \cdot {\mathit f}_{esc}^{Ly\alpha}\cdot (1 - {\mathit f}_{esc}^{LyC})}{4 \cdot \pi \cdot d_{L}^2} \frac{SFR(H\alpha)}{C_{H\alpha}}\cdot 10^{-0.4\cdot A_{Ly\alpha}} 
\end{equation}
where $f\rm_{esc}^{LyC}$ is the escape fraction of ionising photons (i.e. the Lyman continuum photons), $\rm C_{H\alpha}$ is the conversion factor from \cite{Kennicutt+12} (i.e. $\rm C_{H\alpha} = 5.37 \times 10^{-42}~\rm M_\odot~ yr^{-1}~ erg^{-1}~s$) and $\rm A_{Ly\alpha} = E(B-V)\cdot k_{\lambda}(Ly\alpha)$ with $\rm k_{\lambda}(Ly\alpha)$ the \citetalias{Calzetti+00} attenuation law at the \lya\ wavelength.

\begin{figure}
    \centering
    \includegraphics[width = .9\columnwidth]{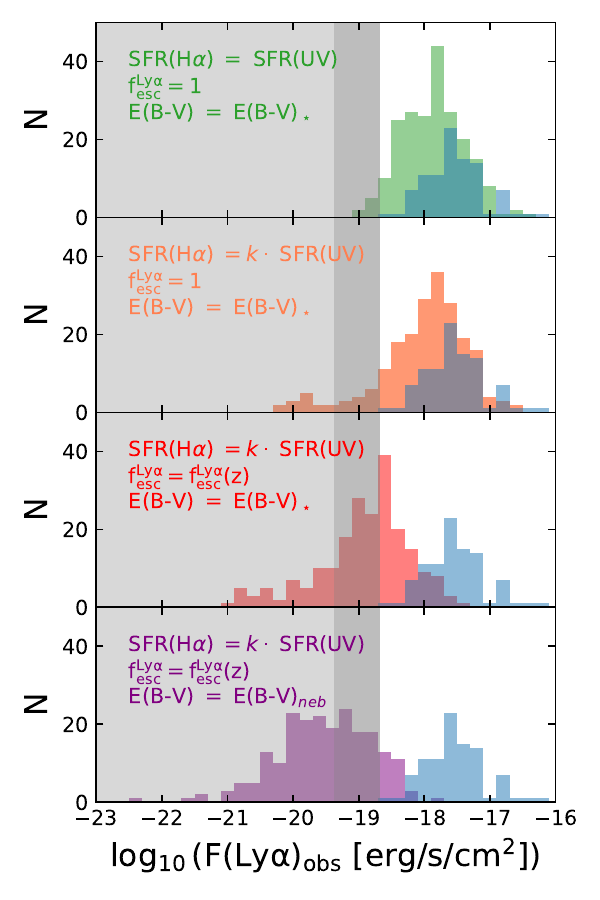}
    \caption{Comparisons between the distribution of the observed \lya\ flux of young ($\leq 10$~Myr) LAEs (in blue) and the expected observed \lya\ flux of young LBGs (in different colours) as derived from Equation~\ref{eq:lya_lbgs} and under different assumptions (reported in the left top corner of each panel). The grey-shaded region is indicative of the fluxes below the $2\sigma$ depth of the available MUSE observations \citep{Bacon+23}, i.e. $2.1 \times 10^{-19}$ and $4.2 \times 10^{-20}$~erg/s/cm$^2$ at 10- and 141-h depths, respectively.}
    \label{fig:expected_lya}
\end{figure}

For our sample of LBGs we do not have direct detection of the H$\alpha$ line. This prevents us from estimating the SFR from the H$\alpha$ luminosity, as well as the Ly$\alpha$ escape fraction. 
Therefore, we first assume $\rm SFR(H\alpha) = SFR(UV)$ and $f\rm_{esc}^{Ly\alpha} = 1$. 
We also set $f\rm_{esc}^{LyC} = 0$. 
With these conditions, we find that $\approx 97$\% of the young LBGs should have an observed \lya\ flux still detectable ($> 2\sigma$) by MUSE.   
This result suggests that, for these sources, the adopted assumptions are not completely valid and other phenomena, e.g. a different SFR estimate, resonant scattering \cite[$ f\rm_{esc}^{Ly\alpha}<1$, e.g.][]{Gronke+16},  and/or dust attenuating the \lya\ in a different way than the UV and optical continuum \cite[\textit{dust selective extinction}, e.g.][]{Neufeld+91, Finkelstein+08, Gronke+16}, should be taken into account.
Hence, as a first step, we investigate the impact of modifying the SFR adopted. 
In the absence of direct spectroscopic detection of H$\alpha$, we attempt to estimate SFR(H$\alpha$) by introducing a correction factor $k$ such as $\rm SFR(H\alpha) = \mathit{k}(age, Z, SFH)\cdot~SFR(UV)$.
The $k$ factor is a function of the stellar population age, metallicity and SFH of each galaxy, i.e. $k = k(\rm age, Z, SFH)$.
We derive $k$ from the \texttt{BPASS} models \cite[version 2.2.1][]{Eldridge+17, Stanway+18} assuming stellar populations with no binary stars, a Chabrier IMF with a cutoff mass of 100 $\rm M_\odot$, a solar and subsolar ($\rm Z = 0.2~Z_\odot$) metallicity and a set of SFH as in our SED modelling, see Section~\ref{subsec:catalogue}. 
For each LBG, we apply to its SFR(UV) the so-derived corresponding k correction factor (see Appendix~\ref{appdx:correction_fact_sfr}). 
The introduction of $\rm k$ reduces the young LBGs with an expected detectable $\rm F(Ly\alpha)_{obs}$ to about 86\%, ~see Figure~\ref{fig:expected_lya}. 

Since correcting the SFR estimate lowers the total number of detectable \lya\ in LBGs by only about 11\%, we investigate the impact of the \lya\ escape fraction on our results. 
To do so, since we do not have a direct estimate of $f\rm_{esc}^{Ly\alpha}$, for each LBG we assume a typical \lya\ escape fraction corresponding to the average $f\rm_{esc}^{Ly\alpha}$ value at the galaxy redshift. 
In particular, we follow the $f\rm_{esc}^{Ly\alpha}$ trend with redshift reported by \cite{Hayes+11}, i.e. $f\rm_{esc}^{Ly\alpha} = 1.67 \cdot 10^{-3} \cdot (1 + z)^{2.57}$.
By doing so, the total number of LBGs with expected observed \lya\ flux above the MUSE detection threshold at $2\sigma$ is about 43\%. 

Finally, we investigate how much the introduction of dust selective extinction could lower the expected observed \lya\ flux of the young LBGs such as that most of the young LBGs would have a \lya\ below the MUSE detection threshold.
In this case, following \cite{Calzetti+94}, we assume that the dust extinction for the emission lines (nebular component) is given by $\rm E(B-V)_{neb} = E(B-V)_\star / 0.58$, valid in the case of the \citetalias{Calzetti+00} attenuation law \cite[][]{Steidel+14}.
This additional condition lowers the number of young LBGs with a detectable \lya\ in MUSE down to only 13\%.

This analysis shows the important role that resonant scattering and (potentially) dust selective extinction have in making the \lya\ undetectable in LBGs.

All in all, these results indicate that LAEs and LBGs are essentially similar \citep[e.g.][]{Dayal+12}. 
We also confirm that resonant scattering and/or dust selective extinction can explain the non-detection of the \lya\ emission in young LBGs.
In contrast to previous studies \cite[e.g.][]{Giavalisco+02, Gawiser+06}, we do not find any evidence that LBGs are higher stellar-mass sources, with less rapid star formation compared to LAEs.

\begin{deluxetable*}{l|ccccccc}[t]
\tablenum{4}
\tablecaption{Table of the median values of the properties of LAEs and mass-matched LBGs.}
\tablewidth{0pt}
\tablehead{
\colhead{} & \colhead{E(B-V)}  & \colhead{M$_\star$} & \colhead{Age} & \colhead{$\beta$} & \colhead{M(UV)} & \colhead{SFR(UV)} & \colhead{sSFR(UV)}\\
\colhead{} & \colhead{}  & \colhead{$\log_{10}$([M$_\odot$])}  & \colhead{$\log_{10}$([yr])} & \colhead{} & \colhead{[mag]} & \colhead{$\log_{10}$([M$_\odot$/yr])} & \colhead{$\log_{10}$([1/yr])}}
\startdata
LAEs & $0.09_{-0.09}^{+0.11}$ & $7.50\pm 0.60$ & $7.17_{-0.57}^{+1.03}$ & $-2.21 \pm 0.56$ & $-18.47_{-1.35}^{+1.25}$  & $-0.05_{-0.50}^{+0.54}$ & $-7.48_{-0.7}^{+0.43}$\\
LBGs & $0.10_{-0.10}^{+0.20}$ & $7.50 \pm 0.51$ & $7.02_{-0.48}^{+1.19}$ &  $-1.95_{-0.54}^{+0.73}$ & $-18.27_{-1.83}^{+1.43}$ & $-0.14_{-0.57}^{+0.73}$ & $-7.46_{-0.94}^{+0.43}$ \\
\enddata
\tablecomments{In the above table, we report in logarithmic value the stellar mass M$_\star$, the  age, the star formation rate SFR(UV) and the specific star formation rate sSFR(UV) and the corresponding 68\% scatter. We also present the values of E(B-V), $\beta$ and dust-corrected M(UV) with their 68\% uncertainties.}
\label{tab:laes_lbgs_prop}
\end{deluxetable*}

\begin{deluxetable*}{l|ccccccc}[t]
\tablenum{5}
\tablecaption{Two-sample KS test between the properties of LAEs and mass-matched LBGs.}
\tablewidth{0pt}
\tablehead{
\colhead{KS test} & \colhead{E(B-V)}  & \colhead{M$_\star$}  & \colhead{Age} & \colhead{$\beta$} & \colhead{M(UV)} & \colhead{SFR(UV)} & \colhead{sSFR(UV)}}
\startdata
$p$-value & $1e-7$ & $0.70_{-0.29}^{+0.23}$ & $0.22_{-0.15}^{+0.23}$ & $1e-4$ & $0.08_{-0.06}^{+0.10}$ & $0.09_{-0.06}^{+0.11}$ & $0.10_{-0.07}^{+0.14}$\\
\enddata
\tablecomments{We report the exact estimate of the $p$-value only for parameters with $p$-value $\geq 0.05$ and for those values with an errorbar that comprises $p$-value $= 0.05$. In all the other cases, we only report its magnitude.}
\label{tab:ks_test_laes_lbgs}
\end{deluxetable*}

\section{Conclusions}\label{sec:conclusions}
In this paper we have studied the physical properties of 182 Lyman-$\alpha$ emitters and 450 stellar-mass matched Lyman-Break galaxies, based on the analysis of their rest-frame UV/optical spectral energy distribution derived from the 28 available filters from \hst\ and \jwst\ in the Hubble XDF.
From this rich multi-wavelength dataset we found that:

\begin{itemize}
    \item Out of 450 initial LAEs in the XDF with a secure spectroscopic identification \citep{Bacon+23}, we did not retrieve a UV/optical counterpart in our photometric catalogue for 72 LAEs (16\%). For the remaining sample, we found that 250 (56\%) match with a single source and 128 (28\%) have multiple sources in their vicinity (up to five possible counterparts within a radius of 0.5 arcsec).
    Similar percentages were also reported by \cite{Bacon+23} who found that 15\% of their sample had no counterpart while 68\% were matched to a single object.
    The addition of further selection criteria to ensure an accurate identification of the LAE counterpart shrunk the sample to 182 LAEs with a single counterpart. 

    \item Based on the study of their photometry, the LAEs in our sample are low-mass systems ($\rm M_\star = 10^{6} - 10^9~M_\odot$) with no or little dust content (E(B-V) = 0 - 0.3) and have blue UV continuum slopes ($\beta = -2.21 \pm 0.56$). The majority of them ($\approx 71$\%) prefers best-fit stellar templates with a sub-solar metallicity. These results are broadly consistent with the past literature \cite[e.g.][and references therein]{Ouchi+20}.
    
    \item The age distribution of our sample of LAEs is bimodal \cite[e.g.][]{Lai+08, Finkelstein+09, Pentericci+09, Rosani+20}. While 75\% of LAEs have an age $< 100$~Myr (young), the remaining 25\% has a significantly older stellar population ($\geq 100$~Myr). A two-sample KS-test on the physical properties of the two samples showed that young and old LAEs have different distributions in E(B-V), stellar mass and sSFR. Specifically, old LAEs are statistically more massive and have lower extinction and sSFR compared to young LAEs. On the contrary, we have not found a statistically significant difference with regard to the overall distributions in \lya\ luminosity, UV continuum slope, UV absolute magnitude and SFR. However, when investigating the regions populated in the M$_\star$ -- SFR plane by these two sub-samples of LAEs, we found that while old LAEs lie along the MS of star-forming galaxies, young LAEs populate the starburst region, displaying a higher SFR at a given stellar mass in comparison to old LAEs.
    This fact hints at the possibility that young and old LAEs are galaxies that went through different evolutionary paths: while young LAEs could be young galaxies undergoing their first burst of star-formation, old LAEs could be old systems experiencing a rejuvenation process triggered by a later sub-halo accretion. 
    From the analysis of the average SED of these two sub-samples, we noticed that \jwst\ observations are crucial to distinguish between these two sub-classes of LAEs in the $z$-range studied. In fact, the rest-frame UV SED of young and old (probed by the \hst\ photometry) is virtually identical.

    \item From our multi-wavelength photometric catalogue of sources in XDF, we found 450 galaxies with photometric redshift in the same $z$-range as our sample of LAEs ($z = 2.8 - 6.7$) and that do not display \lya\ emission. A two-sample KS-test between the overall properties of the LAEs and the mass-matched sample of LBGs did not highlight any statistical difference between these two classes of objects except for the E(B-V) and UV continuum slope. 
    Interestingly, we found that 48\% of the LBGs sample is constituted by objects with a best-fit stellar population age $\leq 10$~Myr. 
    Despite being young, these objects do not display \lya\ emission.
    By looking at their properties, we found that, with respect to young LAEs, young LBGs typically display a higher dust extinction. However, by inferring the expected observed \lya\ flux of these objects from their SFR(UV), we showed that the \lya\ emission of these galaxies should be detectable in the available MUSE observations. This suggests that other mechanisms (e.g. resonant scattering and/or dust-selective extinction) are hampering the detection of the \lya\ emission. 
    All in all, our results indicate that the overall samples of LAEs and LBGs are essentially similar in the main physical properties investigated in this paper \citep[e.g.][]{Dayal+12}, except for higher dust extinction in LBGs, especially at young ages.

\end{itemize}

The results obtained by this study highlight the paramount role of \jwst\ NIRCam and MIRI (deep) imaging surveys in allowing the detection of different classes of Lyman-$\alpha$ emitters, young and old, as well as to unveil the similarities between the stellar population properties of Lyman-$\alpha$ emitters and Lyman-Break galaxies.
The wide wavelength range probed ($0.9 - 5.6~\mu$m) by the available observations, coupled with the high resolution and sensitivity brought by the \jwst\ observatory, allows us to tightly constrain the spectral energy distribution of sources at $z \simeq 3 - 7$ up to their rest-frame optical/near-IR emission in an unprecedented way.
We expect \jwst\ spectroscopic surveys (e.g. NIRCam Wide Field Slitless Spectroscopy) to further push our knowledge of these sources, enabling us to investigate their physical properties in even greater detail.


\section*{Acknowledgments}
The authors thank the anonymous referee for carefully reading the manuscript and providing useful comments.
The authors also thank Josephine Kerutt and Sophie van Mierlo for useful discussions and comments.

EI and KIC acknowledge funding from the Netherlands Research School for Astronomy (NOVA). 
KIC acknowledges funding from the Dutch Research Council (NWO) through the award of the Vici Grant VI.C.212.036.
LC acknowledges financial support from Comunidad de Madrid under Atracci\'on de Talento grant 2018-T2/TIC-11612.
PGP-G acknowledges support from grants PGC2018-093499-B-I00 and PID2022-139567NB-I00 funded by Spanish Ministerio de Ciencia e Innovaci\'on MCIN/AEI/10.13039/501100011033, FEDER, UE.
J.A-M., AC.G. and Luis Colina acknowledge support by grant PIB2021-127718NB-100 from the Spanish Ministry of Science and Innovation/State Agency of Research MCIN/AEI/10.13039/501100011033 and by ``RDF A way of making Europe''. 
AAH acknowledges support from grant PID2021-124665NB-I00 funded by  
MCIN/AEI/10.13039/501100011033 and by ERDF A way of making Europe.
AB acknowledges support from the Swedish National Space Administration (SNSA).
S.E.I.B. acknowledges funding from the European Research Council (ERC) under the European Union’s Horizon 2020 research and innovation programme (grant agreement no. 740246 ``Cosmic Gas'').
A.E. and F.P. acknowledge support through the German Space Agency DLR 50OS1501 and DLR 50OS2001 from 2015 to 2023.
J.H. and D.L. were supported by a VILLUM FONDEN Investigator grant (project number 16599).
TRG and IJ acknowledge support from the Carlsberg Foundation (grant no CF20-0534).
SG acknowledges financial support from the Villum Young Investigator grant 37440 and 13160 and the Cosmic Dawn Center (DAWN), funded by the Danish National Research Foundation (DNRF) under grant No. 140.
The Cosmic Dawn Center (DAWN) is funded by the Danish National Research Foundation (DNRF) under grant No. 140.
JPP and TVT acknowledge financial support from the UK Science and Technology Facilities Council, and the UK Space Agency.
P.-O.L. acknowledges financial support from CNES.
 
This work is based  on observations made with the NASA/ESA/CSA James Webb Space Telescope. The data were obtained from the Mikulski Archive for Space Telescopes at the Space Telescope Science Institute, which is operated by the Association of Universities for Research in Astronomy, Inc., under NASA contract NAS 5-03127 for JWST. These observations are associated with programs GO \#1963, GO \#1895 and GTO \#1283. 
The authors acknowledge the team led by coPIs C. Williams, M. Maseda and S. Tacchella, and PI P. Oesch, for developing their respective observing programs with a zero-exclusive-access period. 
Also based on observations made with the NASA/ESA Hubble Space Telescope obtained from the Space Telescope Science Institute, which is operated by the Association of Universities for Research in Astronomy, Inc., under NASA contract NAS 5–26555.  
The work presented here is the effort of the entire MIRI team and the enthusiasm within the MIRI partnership is a significant factor in its success. MIRI draws on the scientific and technical expertise of the following organisations: Ames Research Center, USA; Airbus Defence and Space, UK; CEA-Irfu, Saclay, France; Centre Spatial de Li\`ege, Belgium; Consejo Superior de Investigaciones Cient\'{\i}ficas, Spain; Carl Zeiss Optronics, Germany; Chalmers University of Technology, Sweden; Danish Space Research Institute, Denmark; Dublin Institute for Advanced Studies, Ireland; European Space Agency, Netherlands; ETCA, Belgium; ETH Zurich, Switzerland; Goddard Space Flight Center, USA; Institute d’Astrophysique Spatiale, France; Instituto Nacional de T\'ecnica Aeroespacial, Spain; Institute for Astronomy, Edinburgh, UK; Jet Propulsion Laboratory, USA; Laboratoire d’Astrophysique de Marseille (LAM), France; Leiden University, Netherlands; Lockheed Advanced Technology Center (USA); NOVA Opt-IR group at Dwingeloo, Netherlands; Northrop Grumman, USA; Max-Planck Institut für Astronomie (MPIA), Heidelberg, Germany; Laboratoire d’Etudes Spatiales et d’Instrumentation en Astrophysique (LESIA), France; Paul Scherrer Institut, Switzerland; Raytheon Vision Systems, USA; RUAG Aerospace, Switzerland; Rutherford Appleton Laboratory (RAL Space), UK; Space Telescope Science Institute, USA; Toegepast-Natuurwetenschappelijk Onderzoek (TNO-TPD), Netherlands; UK Astronomy Technology Centre, UK; University College London, UK; University of Amsterdam, Netherlands; University of Arizona, USA; University of Cardiff, UK; University of Cologne, Germany; University of Ghent; University of Groningen, Netherlands; University of Leicester, UK; University of Leuven, Belgium; University of Stockholm, Sweden; Utah State University, USA

%

\vspace{5mm}
\facilities{{\sl VLT}, {\sl HST}, {\sl JWST}}.

\software{\textsc{Astropy} \citep{astropy:2013, astropy:2018, astropy:2022}, 
\textsc{LePHARE} \citep{Ilbert+06},
\textsc{NumPy} \citep{Harris+20},
\textsc{pandas} \citep{McKinney+10, reback2020pandas}
\textsc{Photutils} \citep{Bradley+22}, 
\textsc{SciPy} \citep{Virtanen+20}
\textsc{Source Extractor} \citep{Bertin+96},
\textsc{SpectRes} \citep{Carnall+17},
\textsc{TOPCAT} \citep{Taylor+11}.
          }
\textsc{BPASS} \citep{Eldridge+17, Stanway+18}




\appendix

\section{Impact of the MIRI/F560W filter}
\label{appdx:f560w_impact}
In the following Appendix, we investigate the impact that the MIRI/F560W filter has in the determination of the physical properties of our sample of 182 LAEs via the SED fitting performed through \lephare.
In particular, we probe possible variations in the estimated values of stellar mass $\rm M_\star$, age and colour excess E(B-V).

To do so, we run \lephare\ on the photometry of our LAEs, after fixing their redshift to the spectroscopic value \citep{Bacon+23}, both considering and excluding the MIRI/F560W filter.
From the comparison of the output best-fit values, we find that 50 (27\%) of our LAEs have a difference in stellar mass larger than 10\%. In particular, the lack of the photometric point at 5.6~$\mu$m tends to underestimate the stellar mass of our galaxies.
Similarly, 62 (34\%) LAEs result to have a difference in stellar age larger than 10\%, with the best-fit SED solutions lacking the F560W filter generally underestimating the age of the underlying galaxy stellar population. 
Finally, we find that 22 (12\%) LAEs have a variation in their estimated E(B-V) larger than 10\%. In this case, the lack of the 5.6~$\mu$m photometry prefers higher values for the colour excess.  
The galaxies with a variation of more than 10\% in all three properties are 12\% (21 objects) of all our sample.

\begin{figure}
    \centering
    \includegraphics[width = .7\textwidth]{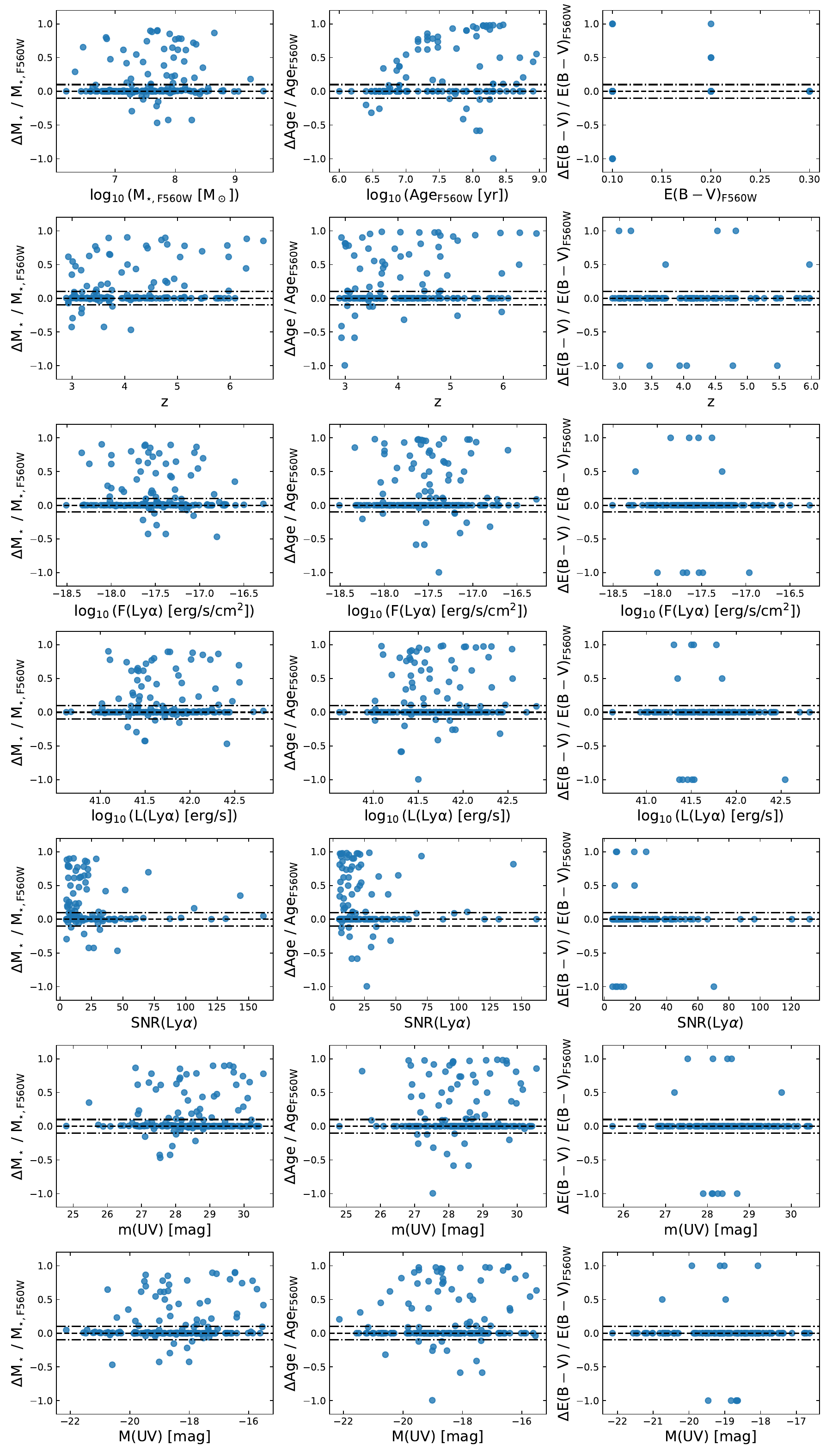}
    \caption{Redshift, \lya\ and UV properties of our sample of LAEs as a function of the variation of their stellar mass, age and colour excess by considering and excluding the MIRI/F560W filter during the SED fitting with \textsc{LePHARE}. The dashed line represents the identity line, while the dash-dotted lines are indicative of 10\% variation. }
    \label{fig:appdx_miri}
\end{figure}

We do not find any evident trend of these discrepancies with redshift nor with the overall \lya\ (e.g. flux, luminosity, SNR) and UV properties (observed and intrinsic UV magnitudes) of the sources, see Figure~\ref{fig:appdx_miri}.

\section{Impact of the SFH}\label{appdx:impact_sfh}
\begin{figure}
    \centering
    \includegraphics[width = \textwidth]{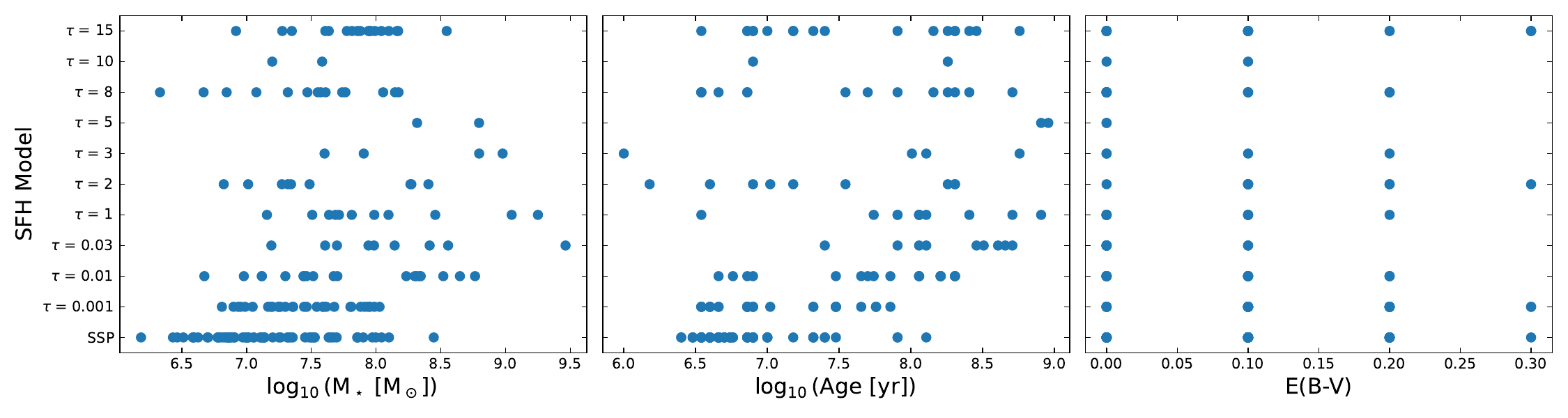}
    \caption{Best fit SED parameters (stellar mass, age and colour excess) as a function of the best fit SED star formation history (instantaneous burst, $\tau$-models).}
    \label{fig:appdx_sfh}
\end{figure}

In the following Appendix, we investigate if the values of the parameters retrieved from the \textsc{LePHARE} SED fitting could be partially driven by the SFH of the best fit.
As reported in Section~\ref{subsec:catalogue}, we allow \textsc{LePHARE} to choose among an  instantaneous burst (i.e. an SSP) and exponentially declining models ($\tau$-models) with ten different values of the $e$-folding time $\tau = 0.001, 0.01, 0.03, 1, 2, 3, 5, 8, 10, 15$~Gyr.
Out of our sample of 182 LAEs, 64 objects (about 35\%) have a best fit SFH reproduced by an SSP while the remaining 118 galaxies (about 65\%) prefer a $\tau$-model. 
In Figure~\ref{fig:appdx_sfh}, we report diagrams showing the values of the main physical parameters retrieved from the SED fitting as a function of the best fit SFH.
We do not recover any correlation that could constitute a bias in the values of the parameters investigated in this study.
We, however, highlight how the stellar mass $\log_{10}(\rm M_\star~ [M_\odot]) \geq 8.5$ and stellar population ages $\log_{10}(\rm Age~ [yr]) > 8.2$ are only reproduced by $\tau$-models.
Interestingly the youngest objects of our sample of LAEs (i.e. $\log_{10}(\rm Age~ [yr]) < 6.3$) are derived for a $\tau$-model with an $e$-folding time of $2-3$~Gyr.

\section{Correction factors SFR(H$\alpha$) - SFR(UV)}
\label{appdx:correction_fact_sfr}
\begin{figure}
    \centering
    \includegraphics[width = .7\textwidth]{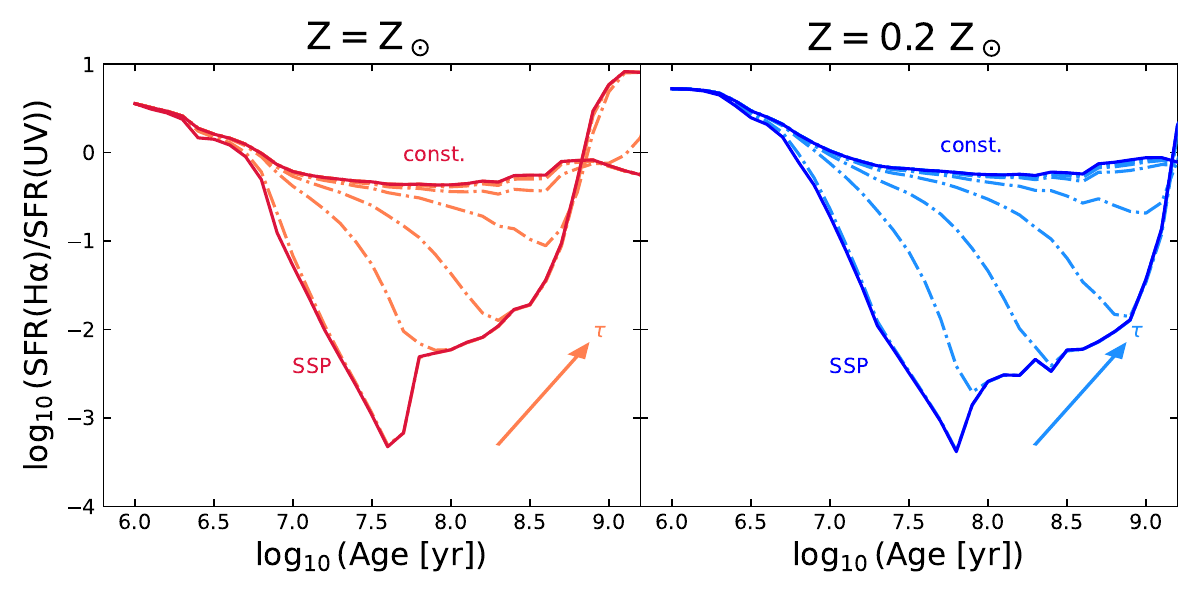}
    \caption{Theoretical tracks for the evolution of the SFR(H$\alpha$) / SFR(UV) ratio as a function of time in the log -- log plane for solar (left panel) and subsolar ($Z = 0.2~Z_\odot$, right panel) metallicities and different SFH: single burst, constant and $\tau$-models ($\tau = 0.001, 0.01, 0.03, 1, 2, 3, 5, 8, 10, 15$~Gyr). 
    The coloured arrow in the bottom right corner of each panel shows the direction of increase of the $e$-folding time $\tau$. }
    \label{fig:appdx_sfr_ratio}
\end{figure}

In Figure~\ref{fig:appdx_sfr_ratio}, we present the theoretical tracks for the evolution of the SFR(H$\alpha$) / SFR(UV) ratio  \cite[assuming the prescriptions by][]{Kennicutt+98} as a function of time, in a log -- log plane, for solar (left panel) and subsolar ($Z = 0.2~Z_\odot$, right panel) metallicities. Different tracks correspond to different SFH. 
We adopt the same SFH models used during the SED fitting of our sources, i.e. single burst and $\tau$-models with $\tau = 0.001, 0.01, 0.03, 1, 2, 3, 5, 8, 10, 15$~Gyr (see Section~\ref{subsec:catalogue}).
To these models, we also add a constant SFH.
The tracks are derived from the \texttt{BPASS} models \citep{Eldridge+17, Stanway+18} for a Chabrier IMF with a cut-off mass of 100 M$_\odot$ and no binary stars.

\bibliography{biblio}{}
\bibliographystyle{aasjournal}



\end{document}